\documentclass[10pt,epsfig]{article}

\usepackage{enumerate}
\usepackage{float}
\usepackage{subfig}
\usepackage{color}
\usepackage[table]{xcolor}
\usepackage{slashed}
\usepackage{multirow}
\usepackage{cite}

\let\counterwithin\relax
\usepackage{chngcntr}
\usepackage{amssymb, amsmath,mathrsfs}

\usepackage{graphics}
\usepackage{graphicx}
\usepackage{epsf}
\usepackage{epsfig}
\usepackage{float}
\usepackage{makecell}
\usepackage{multirow}
\usepackage{color}
\usepackage{xcolor}
\usepackage{simplewick}

\usepackage[utf8]{inputenc}
\usepackage{amsmath}
\usepackage{amssymb}
\usepackage{subfig}
\usepackage[normalem]{ulem}

\usepackage{mathtools}
\usepackage{amsmath}
\usepackage{diagbox}

\newcommand\undermat[2]{
	\makebox[0.5pt][l]{$\smash{\underbrace{\phantom{%
					\begin{matrix}#2\end{matrix}}}_{ \let\scriptstyle\textstyle\text{\large $#1$}}}$}#2}
\newcommand\overmat[2]{
	\makebox[-1pt][l]{$\smash{\overbrace{\phantom{%
					\begin{matrix}#2\end{matrix}}}^{ \let\scriptstyle\textstyle\text{\large $#1$}}}$}#2}    
\usepackage{tikz}
\usepackage[vcentermath]{youngtab}
\usepackage{slashed} 
\usepackage[font=small]{caption} 
\usepackage{times} 

\usepackage{bm}       
\usepackage{bbm} 

\usepackage{relsize}  

\usepackage{autobreak}  

\usepackage{makeidx} 
\usepackage{bbding} 
\usepackage{listings} 
\usepackage{ytableau}

\usepackage[colorlinks=true,
linkcolor=blue,
urlcolor=red,
citecolor=red]{hyperref}

\long\def\rpl#1!!#2!!{\textcolor{red}{#1} \textcolor{blue}{#2}}

\usepackage[top=1in, left=0.95in, bottom=1.1in, right=0.95in]{geometry}
\def\baselinestretch{1.27}
\usepackage[toc]{appendix}

\newcommand{\beq}{\begin {equation}}  
\newcommand{\eeq}{\end   {equation}} 
\newcommand{\bea}{\begin {eqnarray}} 
\newcommand{\eea}{\end   {eqnarray}}  
\newcommand{\baa}{\begin {array}   } 
\newcommand{\eaa}{\end   {array}   }     
\newcommand{\bit}{\begin {itemize} }
\newcommand{\eit}{\end   {itemize} }
\newcommand{\be }{\begin {equation}} 
\newcommand{\ee }{\end   {equation}}

\newcommand{\mc}[1]{\mathcal{#1}}

\newcommand{\bra}[1]{\langle #1 |}
\newcommand{\vev}[1]{ \left\langle {#1}  \right\rangle }

\newcommand{\ie}{{\text{i.e.}}~}

\newcommand{\eq}[1]{\begin{equation}\begin{split} #1 \end{split}\end{equation}}

\newcommand{\4}{\!\!\!\!/\,}
\newcommand{\comment}[1]{}

\newcommand{\TeV}{\ensuremath{\mathrm{TeV}}}

\newcolumntype{M}[1]{>{\centering\arraybackslash}m{#1}}
\newcolumntype{N}{@{}m{0pt}@{}}

\allowdisplaybreaks
\begin{document}

\begin{center}

{\Large \textbf  {Low Energy Effective Field Theory Operator Basis at $d \le 9$}}\\[10mm]

Hao-Lin Li$^{a}$\footnote{lihaolin@itp.ac.cn}, Zhe Ren$^{a, b}$\footnote{renzhe@itp.ac.cn}, Ming-Lei Xiao$^{a}$\footnote{mingleix@itp.ac.cn}, Jiang-Hao Yu$^{a, b, c, d}$\footnote{jhyu@itp.ac.cn}, Yu-Hui Zheng$^{a, b}$\footnote{zhengyuhui@itp.ac.cn}\\[10mm]

\noindent 
$^a${\em \small CAS Key Laboratory of Theoretical Physics, Institute of Theoretical Physics, Chinese Academy of Sciences,    \\ Beijing 100190, P. R. China}  \\
$^b${\em \small School of Physical Sciences, University of Chinese Academy of Sciences,   Beijing 100049, P.R. China}   \\
$^c${\em \small School of Fundamental Physics and Mathematical Sciences, Hangzhou Institute for Advanced Study, University of Chinese Academy of Sciences, Hangzhou 310024, China}\\
$^d${\em \small International Centre for Theoretical Physics Asia-Pacific, Beijing/Hangzhou, China}\\[10mm]

\date{\today}   
          
\end{center}

\begin{abstract} 
    
    We obtain the complete operator bases at mass dimensions 5, 6, 7, 8, 9 for the low energy effective field theory (LEFT), which parametrize  various physics effects between the QCD scale and the electroweak scale. The independence of the operator basis regarding the equation of motion, integration by parts and flavor relations, is guaranteed by our algorithm~\cite{Li:2020gnx,Li:2020xlh}, whose validity for the LEFT with massive fermions involved is proved by a generalization of the amplitude-operator correspondence. At dimension 8 and 9, we list the 35058 (756) and 704584 (3686) operators for three (one) generations of fermions categorized by their baryon and lepton number violations $(\Delta B, \Delta L)$, as these operators are of most phenomenological relevance.

\end{abstract}

\newpage

\setcounter{tocdepth}{4}
\setcounter{secnumdepth}{4}

\tableofcontents

\setcounter{footnote}{0}

\def\baselinestretch{1.5}
\counterwithin{equation}{section}

\newpage

\section{Introduction}

The standard model (SM) has been proved to be a success according to current experiment data at the LHC, yet several unsolved basic problems about our universe motivate both theorists and experimentalists to search for new physics beyond the SM, which have yielded a null result up
to $\Lambda_{\rm NP} \sim \TeV$ scale. Under the circumstances, the effective field theory (EFT) provides a model-independent way to parameterize new physics above $\Lambda_{\rm NP}$ using interactions involving higher dimensional operators at energy scale $<\Lambda_{\rm NP}$.

While the physics above the electroweak scale $\Lambda_{\rm EW}$ is described by the standard model effective field theory (SMEFT)~\cite{Buchmuller:1985jz,Grzadkowski:2010es,Lehman:2014jma,Li:2020gnx,Li:2020xlh,Murphy:2020rsh,Liao:2020jmn}, we also need the low-energy effective field theory (LEFT) below $\Lambda_{\rm EW}$, which has gauge symmetry $SU(3)_C \times U(1)_{\rm EM}$ and all the SM fermions but the top quark.
Part of the LEFT operators can be derived from the SM by integrating out the massive vectors $W^\pm$, $Z$, the top quark $t$ and the Higgs boson $h$. 
For example, the historical Fermi interactions generated by integrating out the $W^\pm$ boson is essentially a part of the LEFT operators at the dimension-6 level, and had been applied to various processes like the $\beta$ decay, the muon decay, and the lepton flavor violation~\cite{Donoghue:1992dd, Ando:2004rk, Falkowski:2020pma, Crivellin:2017rmk}. 
Furthermore, flavor physics such as $B$-physics and $K$-physics utilizes these LEFT operators to parameterize new physics contributions in the $B$ and $K$ meson decay processes, e.g.~\cite{Aebischer:2017gaw,Gonzalez-Alonso:2016etj}. However, generic LEFT operators are only subject to Lorentz invariance and gauge symmetry, which is far more rich than those derived in the SM.

The low energy probes of new physics, complementary to collider searches for new physics, have been gradually gaining importance. Various low energy processes, such as neutrino-less double beta decay~\cite{Prezeau:2003xn, Cirigliano:2018yza}, neutron-antineutron oscillation~\cite{Mohapatra:1980qe,Rao:1982gt,Mohapatra:2009wp,Phillips:2014fgb}, nucleon exotic decays~\cite{Wilczek:1979hc,Abbott:1980zj}, dinucleon decays~\cite{Girmohanta:2019cjm}, electric dipole moment~\cite{Engel:2013lsa,Chang:1992vs}, have been investigated to examine fundamental global symmetries, such as charge conjugate and parity violation, baryon and lepton number violation~\cite{Heeck:2019kgr}. Below the electroweak scale, various new physics effects in these low energy processes could be systematically parameterized in the LEFT. For the baryon and lepton number violation~\footnote{At dimension 5, we find the only violation pattern is $\left(\Delta B, \Delta L\right)=(0,\pm2)$, while there are $(0,\pm2)$, $(0,\pm4)$, $(\pm1,\pm1)$, $(\pm1,\mp1)$ at dimension 6 and $(0,\pm2)$, $(\pm1,\pm1)$, $(\pm1,\mp1)$ at dimension 7. The violation patterns at dimension 8 are the same as that at dimension 6, while there are new violation patterns $(0,\pm6)$, $(\pm1,\pm3)$, $(\pm1,\mp3)$, $(\pm2,0)$ at dimension 9. }, the neutron-anti-neutron oscillation and dinucleon decay process only start to appear at dimension 9 with $\left(\Delta B, \Delta L\right)=(\pm2, 0)$. It is also possible that the dimension 9 operators with $\left(\Delta B, \Delta L\right) = (0, \pm2)$ dominate the neutrino-less double beta decay process~\cite{Cirigliano:2018yza}. 
There are also exotic nucleon decay processes dominated at dimension 9 with $\left(\Delta B, \Delta L\right) = (\pm1,\mp3)$. Therefore, it is necessary to write down the complete set of the higher dimensional LEFT operators. 
Recently~\cite{Jenkins:2017jig, Jenkins:2017dyc, Liao:2020zyx}, the LEFT operators have been written up to dimension 7.

The traditional way to obtain the operator basis is to consider all possible operators, and then eliminate the operator redundancies caused by equation of motion (EOM), integration by parts (IBP), repeated flavor structure and Fierz identities, e.g.~\cite{Grzadkowski:2010es}, which is quite involved when the number of possible structures blows up at higher dimensions.
So far the traditional way has been applied to dimension 5, 6 in Ref.~\cite{Jenkins:2017jig} and dimension 7 in Ref.~\cite{Liao:2020zyx}, which present the result in the form of a set of operators with complicated flavor relations.
On the other hand, the systematical procedure proposed in Ref.~\cite{Li:2020gnx, Li:2020xlh} is capable of generating independent structures of operators automatically, and presents the flavor-specified operators without extra redundancies. 
However, the procedure makes use of the amplitude-operator correspondence~\cite{Ma:2019gtx}, and was designed only for massless amplitudes, which seems to be not compatible with LEFT.
In this work, we present the massive amplitudes in the form introduced in Ref.~\cite{Conde:2016vxs,Arkani-Hamed:2017jhn}, which have a one-to-one correspondence to a set of massless amplitudes when the spin $S\le 1/2$, while the correspondence between massive amplitudes and operators remains valid. 
Hence we argue that the procedure should work as well for EFTs with massive fermions, such as LEFT. 

We sketch the procedure as follows.
We first use the amplitude operator correspondence to convert the problem of finding the operator basis into finding the amplitude basis, which is further divided into separate tasks as to find: 1. a basis of Lorentz factors expressed as functions of spinor-helicity variables; 2. a basis of gauge factors expressed in terms of invariant group tensors. 
The basis of Lorentz factors is generated via the method in ~\cite{Henning:2019enq,Li:2020xlh}, translated from semi-standard Young tableau (SSYT), which after converting to the operators are automatically free from the IBP and EOM redundancy. 
Meanwhile, using the Littlewood-Richardson rules, we construct a set of singlet Young tableau of the gauge group indices from the constituting particles, which induces the basis of gauge group factors. 
A complete and independent flavor-blind operator basis are obtained by direct product of these two factors. 
Afterwards, we symmetrize the flavor indices of the repeated fields to obtain the so-called p-basis operators, where the ones with permutation symmetries allowed by the flavor number make up our final results as an independent basis of flavor-specified operators. 
As a key feature, our form of operators include a flavor Young tableau, which indicates the independent components in the Wilson coefficient tensor, hence no more flavor relations are necessary.

In this work, we apply the procedure above to the LEFT. The dimension 5, 6, 7 operator bases are reproduced with new form of flavor structures as indicated above, compared to the previous literature~\cite{Jenkins:2017jig,Liao:2020zyx}. More importantly, we also generate the dimension 8, 9 operator bases for the first time, which constitutes the main part of this work.
We adopt the chiral fermion notation in order to be consistent with other literature~\cite{Jenkins:2017jig}, as well as to keep the correspondence with the spinor helicity amplitudes explicit.


The paper is organized as follows. In section~\ref{sec:CC}, We introduce the notation and discuss the validity of amplitude-operator correspondence in the massive spinor case. In particular, we describe the main ideas of how to deal with the amplitude bases, group structures and permutation symmetries in section~\ref{sec:OpBa}. In the section~\ref{sec:List}, We list the complete operator bases in order of dimension from 5 to 9. Our conclusion is presented in section~\ref{sec:con}.

\section{General Framework}\label{sec:CC}
In this section, we describe the framework of writing down the operator basis of the LEFT.
We start by reviewing the massive spinor helicity formalism \cite{Conde:2016vxs,Arkani-Hamed:2017jhn}, and then illustrate a one to one correspondence between the massless and massive amplitude basis for particles with spin $S\le1/2$. Together with the massive version of the amplitude operator correspondence that we formulate later, we extend the framework for massless EFTs introduced in \cite{Li:2020gnx,Li:2020xlh} to include massive fermions, and summarize the procedure of enumerating the complete and independent operator basis of LEFT. 
We strongly recommend readers to gain preliminary information in Ref.~\cite{Li:2020gnx,Li:2020xlh} where we elaborate the algorithm for enumerating and constructing of 
operator basis for massless fields, in this work, we only outline the important ingredients of the algorithm in section.~\ref{sec:OpBa}.

\subsection{Amplitude Basis Including Massive Fermions}
Let us first review the massive spinor helicity formalism developed recently in \cite{Arkani-Hamed:2017jhn}, and build the connection between massive spinor and amplitude basis in this subsection. In spinor formalism, a momentum vector can always be decomposed into two spinors, \ie the spinor helicity variables,
\eq{\label{eq:massive_spinor}
	p_{\alpha\dot\alpha}=p_{\mu}\sigma^{\mu}_{\alpha\dot\alpha}=\left(\begin{array}{cc}
		E+p_3 & p_1-ip_2\\ p_1+ip_2 & E-p_3
	\end{array}\right)=\lambda^I_{\alpha}\tilde\lambda_{\dot\alpha I}.
}
where $\alpha,\dot\alpha$ are the $SU(2)_L\times SU(2)_R$ Lorentz indices, $I$ is the little group indices, which is $SU(2)$ for massive particle, and $U(1)$ for massless particle ($I$ can be omitted). Note that $\det p_{\alpha\dot\alpha}=p^{\mu}p_{\mu}=m^2$. Therefore $p_{\alpha\dot\alpha}$ has rank 1 in massless case and rank 2 for massive particles. 
These indices are raised and lowered by the 2-index Levi-Civita symbols defined as, 
\begin{align}
	\epsilon^{12}=-\epsilon^{21}=\epsilon_{21}=-\epsilon_{12}=1,\quad \lambda^{\alpha I}=\epsilon^{\alpha\beta}\lambda^I_{\beta},\quad \lambda_{\alpha I}=\epsilon_{IJ}\lambda^J_{\alpha},\quad \tilde\lambda^{\dot\alpha I}=\epsilon^{\dot{\alpha}\dot{\beta}}\tilde\lambda^I_{\dot\beta},\quad \tilde\lambda_{\dot\alpha I}=\epsilon_{IJ}\tilde\lambda^J_{\dot\alpha}.
\end{align}
We normalize the spinors as
\begin{align}
	\lambda^{I\alpha}\lambda^J_{\alpha}=-m\epsilon^{IJ},\quad \tilde{\lambda}^I_{\dot\alpha}\tilde{\lambda}^{J\dot\alpha}=m\epsilon^{IJ},\quad 
	\lambda^I_{\alpha}\lambda_{\beta I}=m\epsilon_{\alpha\beta},\quad \tilde{\lambda}^I_{\dot\alpha}\tilde{\lambda}_{\dot\beta I}=m\epsilon_{\dot\alpha\dot\beta} \label{eq:spin_sum}
\end{align}
In the standard polar coordinates where $p^\mu=(E,p\sin\theta\cos\phi,p\sin\theta\sin\phi,p\cos\theta)$, the general solutions of the spinors are the following,
\begin{align}
	\lambda^I_{\alpha}=\lambda_{\alpha}\zeta^{-I}+\eta_{\alpha}\zeta^{+I},\;\text{ with }\lambda_{\alpha}=\sqrt{E+p}\left(\begin{array}{c}
		-s_{\theta}^*\\c_{\theta}		
	\end{array}\right),\;\eta_{\alpha}=\sqrt{E-p}\left(\begin{array}{c}
		c_{\theta}\\s_{\theta}
	\end{array}\right),\label{eq:lambda}\\
	\tilde\lambda^I_{\dot\alpha}=\tilde\lambda_{\dot\alpha}\zeta^{+I}-\tilde\eta_{\dot\alpha}\zeta^{-I},\;\text{ with }\tilde\lambda^{\dot\alpha}=\sqrt{E+p}\left(\begin{array}{c}
		c_{\theta}\\s_{\theta}		
	\end{array}\right),\;\tilde\eta^{\dot\alpha}=\sqrt{E-p}\left(\begin{array}{c}
		s^*_{\theta}\\-c_{\theta}		
	\end{array}\right).\label{eq:lambdatilde}
\end{align}
where $\zeta^{\pm}$ satisfy $\epsilon_{IJ}\zeta^{-I}\zeta^{+J}=1$ and abbreviation $c_\theta \equiv \cos\frac\theta2$, $s_\theta \equiv \sin\frac\theta2 e^{i\phi}$. $\zeta^{\pm}$ are determined by the direction along which we take the spin components, for instance $\zeta^+=(1,0),\;\zeta^-=(0,1)$ mean that we take the spin components along the momentum direction. 
Because we work in the in-coming convention in this paper, the out-going physical momenta are thus written as in-coming momentum with $E < 0$. For such momentum, we define $\lambda^I_{\alpha}(-p)=-\lambda^I_{\alpha}(p)$ and $\tilde{\lambda}^I_{\dot\alpha}(-p)=\tilde{\lambda}^I_{\dot\alpha}(p)$. 
In the massless limit, $E-p,\eta,\tilde\eta\rightarrow 0$, the $SU(2)$ little group reduces to the transverse $U(1)$ rotation, while $I$ can be omitted. Only two of spinors in eq.~(\ref{eq:lambda},\ref{eq:lambdatilde}) are still valid,
\begin{align}
	\lambda_\alpha = \sqrt{2E}\begin{pmatrix} -s_\theta^* \\ c_\theta \end{pmatrix}, \quad 
        \tilde\lambda^{\dot\alpha} = \sqrt{2E}\begin{pmatrix} c_\theta \\ s_\theta \end{pmatrix}. 
\end{align}
The resulting Lorentz scalar can be represented by angle spinor brackets and square spinor brackets
\begin{align}
	\langle 1^I2^J\rangle\equiv\lambda^{\alpha I}_1\lambda^J_{2\alpha}=\lambda^I_{1\alpha}\epsilon^{\beta\alpha}\lambda^J_{2\beta}=-\lambda^I_{1\alpha}\epsilon^{\alpha\beta}\lambda^J_{2\beta}=-\langle 2^J1^I\rangle,\\
	[1^I2^J]\equiv\tilde\lambda^I_{1\dot\alpha}\tilde\lambda^{\dot\alpha J}_2=\tilde\lambda^I_{1\dot\alpha}\epsilon^{\dot\alpha\dot\beta}\tilde\lambda^J_{2\dot\beta}=-\tilde\lambda^I_{1\dot\alpha}\epsilon^{\dot\beta\dot\alpha}\tilde\lambda^J_{2\dot\beta}=-[2^J1^I].
\end{align}
The same definitions apply to the massless case with the $I,J$ indices removed. The Lorentz invariant amplitude can be constructed from the spinor brackets. 
The little group representations of each particle constrains the form of amplitude as follows
\begin{align}
	&\text{ helicity }h\text{ massless particle with spinor variables }(\lambda,\tilde\lambda):\quad \mathcal{A} \sim \left\{\begin{array}{ll} \lambda^{r-2h}_{\{\alpha\}}\tilde{\lambda}_{\{\dot\alpha\}}^{r},\quad h\leq 0 \\ 
	\lambda^{r}_{\{\alpha\}}\tilde{\lambda}_{\{\dot\alpha\}}^{r+2h},\quad h\geq 0
	\end{array}\right.\label{eq:LGless} \\
	&\text{ spin }S\text{ massive particle with spinor variables }(\lambda^I,\tilde\lambda^I):\quad \mathcal{A} \sim \left(\lambda^{r+2S-n}_{\{\alpha\}}\tilde{\lambda}^{r+n}_{\{\dot\alpha\}}\right)^{\{I_1\cdots I_{2S}\}},\quad 0\le n\le 2S \label{eq:LGive}
\end{align}
where $\{\cdot\}$ denotes totally symmetric indices. The total symmetries among the spinor indices are superfluous for the massless amplitudes, while those for massive amplitudes result from the removal of terms with factor of masses via eq.~\eqref{eq:spin_sum}. For example
\eq{\label{eq:ampeg}
	\langle 1^J2\rangle\langle 1^I3\rangle[1_J4] &= \left[\left(\lambda_1^{\{\alpha}\lambda_1^{\beta\}}\tilde\lambda_{1\dot\alpha}\right)^{I} + \underbrace{\left(\lambda_1^{[\alpha}\lambda_1^{\beta]}\tilde\lambda_{1\dot\alpha}\right)^{I}}_{=m_1\epsilon^{\alpha\beta}\tilde\lambda_{1\dot\alpha}^I}\right]\lambda_{2,\alpha}\lambda_{3,\beta}\tilde\lambda_4^{\dot\alpha}.
}
By this rule, we have a well-defined dimensionality of the amplitudes as the highest power of spinor brackets in it. Under this definition, the complete amplitude basis at a certain dimension is defined modulo lower dimensional amplitudes. 
It will be clear shortly that the EOM redundancy of the corresponding operators is eliminated through this removal.
Also due to the total symmetry of the spinor indices, the $r$ pairs of little group indices in eq.~\eqref{eq:LGive} must be contracted between $\lambda^I$ and $\tilde\lambda^I$, because otherwise the amplitude would vanish.
Therefore the integer $r$ stands for the number of spinor pairs $(\lambda^{(I)},\tilde\lambda^{(I)})$ in both cases that constitute factors of momentum due to eq.~\eqref{eq:massive_spinor}.
The $2S+1$ choices of $n$ represent independent forms\footnote{It is true that $\lambda^I$ and $\tilde\lambda^I$ are not independent due to the Dirac equations. However, given the standard form in eq.~\eqref{eq:LGive}, by removing the terms with factor of masses, the amplitude with various $n$ are indeed independent.}
of contribution the spin-$S$ massive particle could have.

It is easy to observe the correspondence between the massless factor and the massive factor when $S=|h|$ and $n=0,2S$, which maps momenta to momenta, and the rest of the massless spinors to massive spinors with totally symmetric little group indices\footnote{
The mapping is referred to as the ``first reduction'' in \cite{Durieux:2020gip}, which is claimed to be valid only for the ``maximal helicity categories'', namely $n = 0,2S$. Our formulation is from a different perspective, which better fits in our framework.
}. 
The mapping covers all the possible massive amplitudes only when $S\le \frac12$, in that $n=0,2S$ are the only two choices. 
Therefore, we have the explicit form of the mapping as
\eq{\label{eq:map2mass}\begin{array}{lcr}
	h=0             &\lambda^{r}\tilde{\lambda}^{r}\stackrel{}{\longrightarrow} \left(\lambda^J\tilde{\lambda}_J\right)^r                   & S=0,\ n=0\\
	h=-\frac12      &\lambda^{1+r}\tilde{\lambda}^r\stackrel{}{\longrightarrow} \lambda^I\left(\lambda^J\tilde{\lambda}_J\right)^r          & S=\frac12,\ n=0\\
	h=\frac12       &\lambda^{r}\tilde{\lambda}^{1+r}\stackrel{}{\longrightarrow} \tilde{\lambda}^I\left(\lambda^J\tilde{\lambda}_J\right)^r  & S=\frac12,\ n=1
\end{array}}
For higher spin massive particles, the map could not be established, and our framework would fail.
However, it is already enough for us to enumerate the amplitude basis involving massive scalars and fermions by enumerating the massless ones and then simply adding the little group indices according to eq.~\eqref{eq:map2mass}.

\subsection{Amplitude-operator correspondence including massive fermions}\label{sec:massless}
With the massive local amplitude basis in hand, we still need the massive version of the amplitude operator correspondence to finally obtain the operator basis. First we convert the Lorentz indices $\mu$ in the operators into the spinor indices $\alpha$ and $\dot\alpha$ and we have the following definitions of various notations used in constructing operator basis:  
\begin{align}
	\gamma^{\mu}=\left(\begin{array}{cc}
		0&\sigma^{\mu}_{\alpha\dot\alpha}\\\bar{\sigma}^{\mu\dot\alpha\alpha}&0
	\end{array}\right),&\quad \Psi=\left(\begin{array}{c}
		\xi_{\alpha}\\\chi^{\dagger\dot\alpha}
	\end{array}\right),\quad \Psi_M=\left(\begin{array}{c}
		\zeta_{\alpha}\\\zeta^{\dagger\dot\alpha}
	\end{array}\right), \label{eq:fermionfield}\\
	\Psi_{\rm L}=\frac{1-\gamma_5}{2}\Psi=\left(\begin{array}{c}
		\xi_{\alpha}\\ 0
	\end{array}\right)&,\quad \Psi_{\rm R}=\frac{1+\gamma_5}{2}\Psi=\left(\begin{array}{c}
		0\\\chi^{\dagger\dot\alpha}
	\end{array}\right),\\
	D^{\mu}=\frac12 D_{\alpha\dot\alpha}\bar{\sigma}^{\mu\dot\alpha\alpha},\quad F_{\rm{L}\mu\nu}&=\frac14 F_{\rm{L}\alpha\beta}\epsilon_{\dot\alpha\dot\beta}\bar{\sigma}^{\mu\dot\alpha\alpha}\bar{\sigma}^{\nu\dot\beta\beta},\quad F_{\rm{R}\mu\nu}=\frac14 F_{\rm{R}\dot\alpha\dot\beta}\epsilon_{\alpha\beta}\bar{\sigma}^{\mu\dot\alpha\alpha}\bar{\sigma}^{\nu\dot\beta\beta},
\end{align}
where $\xi,\chi,\zeta$ are the Weyl fermions, $\Psi$ and $\Psi_M$ denote the Dirac and Majorana fermion respectively, $F_{\rm{L}/\rm{R}}=\frac12 (F\pm i\tilde{F}) $ are the chiral basis of the gauge bosons. 
Note that in terms of Weyl components, we need a pair of them to represent the independent chiral components of the Dirac fermion, but only one for a Majorana fermion. Nevertheless, all the Weyl degrees of freedom share the same form of spinor wave functions
\begin{align}
	u^I=\left(\begin{array}{c}
		\lambda_{\alpha }^I\\\tilde{\lambda}^{\dot\alpha I}
	\end{array}\right),\;\bar{u}_I=(-\lambda^{\alpha }_I,\;\tilde{\lambda}_{\dot\alpha I}),\quad v_I=\left(\begin{array}{c}-\lambda_{\alpha I}\\ \tilde{\lambda}^{\dot{\alpha}}_I\end{array}\right),\quad \bar{v}^I=\left(\lambda^{\alpha I},\;\tilde{\lambda}_{\dot{\alpha}}^I \right),
\end{align}
which solve the Dirac equations
\begin{align}
	(p\!\!\!/-m)u^I=0,\quad\bar{u}_I(p\!\!\!/-m)=0,\quad \bar{u}_Iu^J=2m\delta_I^J,\quad u^I\bar{u}_I=p\!\!\!/+m,\\
	(p\!\!\!/+m)v_I=0,\quad\bar{v}^I(p\!\!\!/+m)=0,\quad \bar{v}^Iv_J=2m\delta^I_J,\quad v_I\bar{v}^I=p\!\!\!/-m.
\end{align}
We focus on $u^I$ and $\bar{v}^I$ associated with incoming fermions and anti-fermions, since all momenta are incoming in our convention. 
The amplitudes generated by fermion bilinears are given by the following Feynman rules.
\begin{align}
	\bar\Psi _1\Psi_2=&\bar\Psi_{{\rm L}1}\Psi_{{\rm R}2}+\bar\Psi_{{\rm R}1}\Psi_{{\rm L}2}\rightarrow\; \bar{v}_1^Iu^J_2=\lambda^{\alpha I}_1\lambda^J_{2\alpha}+\tilde\lambda^I_{1\dot\alpha}\tilde\lambda^{\dot\alpha J}_2=\langle 1^I2^J\rangle+[1^I2^J],\\
	\bar\Psi _1\gamma_5\Psi_2=&\bar\Psi_{{\rm L}1}\Psi_{{\rm R}2}-\bar\Psi_{{\rm R}1}\Psi_{{\rm L}2}\rightarrow\;  \bar{v}_1^I\gamma_5 u^J_2=[1^I2^J]-\langle 1^I2^J\rangle,\\
	\bar\Psi _1\gamma^{\mu}\Psi_2=&\bar\Psi_{{\rm R}1}\gamma^{\mu}\Psi_{{\rm R}2}+\bar\Psi_{{\rm L}1}\gamma^{\mu}\Psi_{{\rm L}2}\rightarrow\; \bar{v}_1^I\gamma^{\mu}u^J_2= \langle 1^I|\sigma^{\mu}|2^J]+[1^I|\bar\sigma ^{\mu}|2^J\rangle,\\
	\bar\Psi _1\gamma^{\mu}\gamma_5\Psi_2=&\bar\Psi_{{\rm R}1}\gamma^{\mu}\Psi_{{\rm R}2}-\bar\Psi_{{\rm L}1}\gamma^{\mu}\Psi_{{\rm L}2}\rightarrow\; \bar{v}_1^I\gamma^{\mu}\gamma_5u^J_2= \langle 1^I|\sigma^{\mu}|2^J]-[1^I|\bar\sigma ^{\mu}|2^J\rangle.
\end{align}
Therefore the amplitude operator correspondence between spinor variables with free little group indices and four-component spinor fields can be read from the second and the last expressions in above equations:
\begin{align}
	\lambda_i^I \rightarrow \psi_i =\Psi_{{\rm L} i},\bar\Psi_{{\rm R} i}\text{ or }\Psi_{M i},\quad \tilde{\lambda}^I_i\rightarrow \psi_i^{\dagger} =\bar{\Psi}_{{\rm L} i},\Psi_{{\rm R} i}\text{ or }\bar{\Psi}_{M i}.
\end{align}
The pair of spinor helicity variables $\lambda^J_i\tilde\lambda_{iJ}$ with contracted little group indices can be translated into derivative acting on particle $i$ yielding following correspondences: 
\begin{align}\label{eq:derivative1}
	\lambda_i^{n+2}\tilde{\lambda}_i^n \Leftrightarrow D^n F_{{\rm L}i},&\quad \lambda_i^n\tilde\lambda^{n+2} \Leftrightarrow D^n F_{{\rm R}i},\\\label{eq:derivative2}
	\lambda_i^{n+1}\tilde{\lambda}_i^n,\;\lambda_i^I\left(\lambda_i^J\tilde\lambda_{iJ}\right)^n \Leftrightarrow D^n\psi_i,&\quad \lambda_i^n\tilde{\lambda}^{n+1}_i,\;\tilde\lambda_i^I\left(\lambda_i^J\tilde\lambda_{iJ}\right)^n \Leftrightarrow D^n\psi_i^{\dagger}.
\end{align}
where for completeness we also include the massless case denoted by spinor variables without $I$ or $J$ indices.
One may question about do the presence of the covariant derivatives generate other local amplitudes with more gauge bosons, which exist in ordinary Feynman rules.  However, these vertices are not gauge invariant, and the final gauge invariant amplitudes with contributions from these operators are non-local, and our amplitude operator correspondence applies to local amplitudes only.

Let us take a closer look at the derivatives in eq.~\eqref{eq:derivative1} and~\eqref{eq:derivative2}. The total symmetries among the spinor indices in amplitudes are very clear. Otherwise, the resulting amplitude must vanish or reduce the dimension due to the on-shell condition $\lambda^I_{i[\alpha}\lambda^J_{i\beta]} = m_i\epsilon^{IJ}\epsilon_{\alpha\beta}$. Therefore the derivatives acting on each fields in the operators also take the totally symmetric spinor indices. It is easy to see that any pair of anti-symmetric spinor indices if associated with derivatives can be always converted to other types or reduced dimesion with EOM and the relation $i[D_\mu,D_\nu] = \sum F_{\mu\nu}$, where we are supposed to have obtained a complete basis:
\eq{
	& D_{[\alpha\dot\alpha}D_{\beta]\dot\beta} = D_{\mu}D_{\nu}\sigma^{\mu}_{[\alpha\dot\alpha}\sigma^{\nu}_{\beta]\dot\beta} = -{\color{red} D^2 }\epsilon_{\alpha\beta}\epsilon_{\dot\alpha\dot\beta} + \frac{i}{2}[D_{\mu},D_{\nu}]\epsilon_{\alpha\beta}\bar\sigma^{\mu\nu}_{\dot\alpha\dot\beta}, \\
	& D_{[\alpha\dot\alpha}\psi_{\beta]} = D_{\mu}\sigma^{\mu}_{[\alpha\dot\alpha}\psi_{\beta]} = -\epsilon_{\alpha\beta}{\color{red} (D\4\psi)_{\dot\alpha} }, \\
	& D_{[\alpha\dot\alpha}F_{{\rm L} \beta]\gamma} = D_{\mu}F_{\nu\rho} \sigma^{\mu}_{[\alpha\dot\alpha}\sigma^{\nu\rho}_{\beta]\gamma} = 2{\color{red} D^{\mu}F_{\mu\nu} } \epsilon_{\alpha\beta}\sigma^{\nu}_{\gamma\dot\alpha}.
}
Therefore, the building block in constructing the operator basis is in the following form:
\begin{equation}
    \left(D^{r_i-|h_i|}\Phi_{i}\right)^{\dot{\alpha}_i^{r_i+h_i}}_{\alpha_i^{r_i-h_i}}\in (j_l+r_i-|h_i|, j_r+r_i-|h_i|)
\end{equation}
where the powers of the indices $\alpha_i$ and $\dot{\alpha}_i$ indicate that they are already totally symmetrized, thus the whole object transforms as $(j_l+r_i-|h_i|{2}, j_r+r_i-|h_i|)$ under $sl(2,\mathbb{C})=su(2)_l\times su(2)_r$, given that $\Phi$ transforms as $(j_l, j_r)$.

The IBP redundancy of operators is solved by the manifest momentum conservation in the amplitude, which can be illustrated by the following example:
$$\vev{1^I2}[23^J] = \bra{1^I}p_2|3^J] = -\bra{1^I}(p_1+p_3+p_4)|3^J] \sim -\vev{1^I4}[43^J] $$
corresponds to the operator equivalence
$$(\psi_1\sigma^\mu\psi^\dagger_3)D_\mu \phi_2 \phi_4 \sim -(\psi_1\sigma^\mu\psi^\dagger_3) \phi_2 D_\mu\phi_4,$$
where terms that convert to other type steming from $\vev{1^I1^J}=m_1\epsilon^{IJ}$ s by EOM are omitted. Hence, taking momentum conservation into account, the amplitude basis corresponds to an IBP non-redundant basis of operators.

Finally, taking into account the $SU(3)$ group factors $T^{a_1,...a_N}$, the amplitude-operator correspondence can be represented by:
\begin{eqnarray}\label{eq:correspond}
&& T^{a_1,...,a_N}\mc{B}(\lambda_1^{(I)},...,\lambda_N^{(I)})\sim T^{a_1,...,a_N}\mathcal{B}\left(\Phi_1,\cdots, \Phi_N\right)_{a_1,\cdots,a_N}\\
	&& \mathcal{B}\left(\Phi_1,\cdots, \Phi_N\right)_{a_1,\cdots,a_N}=\left(\epsilon^{\alpha_i\alpha_j}\right)^{\otimes n}\left(\tilde\epsilon_{\dot{\alpha}_i\dot{\alpha}_j}\right)^{\otimes\tilde{n}}\prod^N_{i=1}\left(D^{r_i-|h_i|}\Phi_{i,a_i}\right)^{\dot{\alpha}_i^{r_i+h_i}}_{\alpha_i^{r_i-h_i}},\label{eq:amp} 
\end{eqnarray}
where $\mc{B}$ and $T$ determine the Lorentz structure and gauge structure of the operators respectively, and our goal to find complete and independent operator basis is equivalent to find independent $T$'s and $B$'s given the type of the operators. Note that we have suppressed the flavor indices on the right hand side of eq.~\eqref{eq:correspond} for the moment , and thus call them as ``flavor-blind" operators, where we effectively treat all the particles as distinguishable ones, the redundancy related to the repeated fields or equivalently constraints from spin-statistics among the identical particles are tackled in the next section.

\subsection{The Operator Basis}\label{sec:OpBa}

Having established the correspondence, the next step is to construct concrete amplitude basis satisfying momentum conservation and spin-statistics constraint. Such amplitude basis could be translated into operator basis following eq.~\eqref{eq:correspond}.

We start by reviewing the operator classification at different levels as introduced in Ref.~\cite{Li:2020gnx, Li:2020xlh}: 
\begin{itemize}
    \item Class: A (Lorentz) class is defined by the numbers of fields of each Lorentz irreducible representations and the definite number of covariant derivatives in a operator. 
    
    \item Type: For each Lorentz Class, an assignment of each Lorentz irreducible representations with SM field contents that conserve the electric charge is called a type.
    
    \item Term (flavor-blind operator): For each type, we organize operators into different irreducible representations of the symmetric groups of permuting flavor indices of repeated fields. Each irreducible representation corresponds to a term of operators, and our result in the following sections are all presented at this level.
    
    \item Operator (flavor-specified operator): For each term, when the flavor indices are taken for a specific number in a SSYT of the corresponding Young diagram, it becomes what we called flavor-specified operator.
\end{itemize}

Since our results are presented at the term level, in the following we briefly summarize the method for finding terms for each type of operators, which has been elaborated in Ref.~\cite{Li:2020gnx, Li:2020xlh}. We encourage readers to refer Ref.~\cite{Li:2020gnx, Li:2020xlh} for details and the systematic explanation of the method. 
In the following, we first separate the task for finding complete and independent Lorentz and gauge factors, then illustrate how to obtain the term with definite flavor permutation symmetry.

In general, the Lorentz factor of the amplitude, determined only by the helicities of the constituting particles, can be expressed as
\eq{\label{eq:amplitude_basis}
	\mathcal{B} = \vev{\cdot}^n [\cdot]^{\tilde{n}},
}
where $n$ and $\tilde{n}$ denote the numbers of $\lambda$ and $\tilde{\lambda}$ pairs, as shown in eq.~(\ref{eq:amp}).

To find the complete and independent amplitude basis, we use the method introduced by Ref.~\cite{Henning:2019enq} and further developed by Ref.~\cite{Li:2020gnx}. It is proved in Ref.~\cite{Henning:2019enq} that the amplitudes modulo total momenta form irreducible representations of $SU(N)$ group are denoted by the primary Young diagrams,
\begin{eqnarray}\label{eq:primary_YD}
	Y_{N,n,\tilde{n}} \quad = \quad \arraycolsep=0pt\def\arraystretch{1}
	\rotatebox[]{90}{\text{$N-2$}} \left\{
	\begin{array}{cccccc}
		\yng(1,1) &\ \ldots{}&\ \yng(1,1)& \overmat{n}{\yng(1,1)&\ \ldots{}\  &\yng(1,1)} \\
		\vdotswithin{}& & \vdotswithin{}&&&\\
		\undermat{\tilde{n}}{\yng(1,1)\ &\ldots{}&\ \yng(1,1)} &&&
	\end{array}
	\right.
	\\
	\nonumber 
\end{eqnarray}
where $N$ is the number of particles in the amplitudes. The base vectors of the irreducible representations are given by $SU(N)$ semi-standard Young tableaus (SSYTs). The number of indices $i$ to fill in the Young diagram, denoted by $\#i$, is determined by $\#i=\tilde{n}-2h_i$, where $\{h_i\}$ denote the set of helicity of the $i$th particle in the class and are sorted in the order $h_i \leq h_{i+1}$, $i=1,\cdots,N-1$. Going through all possible ways to fill in the numbers and obtain a SSYT will give us the complete and non-redundant amplitude basis expressed by SSYTs. In fact, the Fock's condition of Young tableaus corresponds to IBP and Fierz/Schouten identities of operators. Thus choosing SSYTs means picking out all independent Young tableaus, which corresponds to picking out all independent Lorentz structures of operators. 

The SSYTs can be translated to amplitudes using following relations for each columns,
\eq{
	\young(i,j) \sim \vev{ij}, \qquad \begin{array}{c} \young({{k_1}},{{k_2}}) \\ \vdots \\ \young({{k_{N-3}}},{{k_{N-2}}}) \end{array} \sim \mc{E}^{k_1\dots k_{N-2}ij}[ij],
}
where $\mc{E}$ is the Levi-Civita tensor of the $SU(N)$ group. These amplitudes can be further translated to Lorentz structures of operators using amplitude-operator correspondence.

The gauge group factors are expressed by Levi-Civita tensors that contract with the fundamental
indices of the fields. Anti-fundamental representation and adjoint representation of fields can be converted to fundamental representation by
\eq{
	&\epsilon_{acd}\left(\lambda^A\right){}_b^d G^A = G_{abc} \sim \young(ab,c) \ ,\\
	&\epsilon_{abc}u^{\dagger,{c}} = u^\dagger_{ab} \sim \young(a,b) \ .
}
Following the procedures in Ref.~\cite{Li:2020gnx}, the gauge group factors can be obtained by constructing the singlet Young tableau using the modified Littlewood-Richardson rule with the corresponding indices of each field filled in.

Finally, we need to find the flavor structure in the presence of repeated fields. To achieve that, we utilize the basis $b^{[\lambda]}_{x=1,\dots,d_\lambda}$ in the irreducible left ideal of the group algebra $\tilde{S}_m$, which is a set of independent symmetrizers spanning an irreducible representation space of the permutation group $S_m$. By acting them on the flavor indices of an operator, we obtain either zero or an irreducible flavor tensor, whose independent components are given by the $SU(n_f)$ SSYT's. We thus express the result as such irreducible tensors in the form $\mc{Y}^{[\lambda]}_1\circ\mc{O}$, where $\mc{Y}^{[\lambda]}_1\equiv b^{[\lambda]}_1$ is the famous Young symmetrizer\footnote{In our prescription, $\mc{Y}^{[\lambda]}_{i\neq1}\circ\mc{O}$, if not vanishing, belongs to another irreducible tensor $\mc{Y}^{[\lambda]}_1\circ\mc{O}'$, and should not be counted.} and $\mc{O}$ is some monomial operator. 
Take $S_3$ group for example, we show the action of Young symmetrizer as following
\eq{
	&\mc{Y}^{[3]}_1 O^{prs} = \mc{Y}\left[\tiny{\young(prs)}\right] O^{prs} = O^{prs}+O^{rps}+O^{psr}+O^{rsp}+O^{srp}+O^{spr}, \\
	&\mc{Y}^{[2,1]}_1 O^{prs} = \mc{Y}\left[\tiny{\young(pr,s)}\right] O^{prs} = O^{prs}+O^{rps}-O^{srp}-O^{spr}, \\
	&\mc{Y}^{[1^3]}_1 O^{prs} = \mc{Y}\left[\tiny{\young(p,r,s)}\right] O^{prs} = O^{prs}-O^{rps}-O^{psr}+O^{rsp}-O^{srp}+O^{spr}.
}
To get all the independent irreducible representation spaces, one must find a collection of operators $\{\mc{O}_\zeta\}$ such that $\mc{Y}^{[\lambda]}_1\circ\mc{O}_\zeta$ are linearly independent and whose linear combinations exhaust all the possible $\mc{Y}^{[\lambda]}_1\circ\mc{O}$. Such projections form the so-called p-basis of the operators \cite{Li:2020xlh}. 
It is usually highly non-trivial considering the complexity of redundancy relations of the operators. But with the Young tableau basis obtained above, we could find the unique coordinates of any $\mc{Y}^{[\lambda]}_1\circ\mc{O}_\zeta$ which makes it easy to find the collection $\{\mc{O}_\zeta\}$.

In practice, the coordinates of $\mc{Y}^{[\lambda]}_1\circ\mc{O}_\zeta$ can be obtained by finding the matrix representation of the group elements $\mc{D}^{T}(\pi)$ and $\mc{D}^{\mc{B}}(\pi)$ on the complete and independent bases of $T_i$ and $\mc{B}_j$ obtained by the Young tableau approaches discussed above.
As the permutation on the flavor indices is equivalent to permute the gauge and Lorentz structures as discussed in Ref.~\cite{Li:2020gnx,Li:2020xlh}:
\begin{eqnarray}
\underbrace{\pi\circ {\cal O}^{\{f_{k},...\}}}_{\rm permute\ flavor} &=& \underbrace{\left(\pi\circ T_{{\rm SU3}}^{\{g_k,...\}}\right)}_{\rm permute\ gauge}\underbrace{\left(\pi\circ \mc{B}^{\{f_k,...\}}_{\{g_{k},...\}}\right)}_{\rm permute\ Lorentz},
\label{eq:tperm}
\end{eqnarray}
Thus for any operator $\mc{O}=C_{ij}T_iB_j$, the permutation yields a transformation of the coordinate $C_{ij}$:
\begin{eqnarray}
\pi\circ \mc{O}&=&C_{ij}\left(\pi\circ T_i\right)\left(\pi\circ B_j\right)\nonumber \\
&=&C_{ij}\left(\mc{D}^{T}(\pi)_{ik}T_k\right)\left(\mc{D}^{\mc{B}}(\pi)_{il}\mc{B}_l\right)\nonumber \\
&=&\left(C_{kl}\mc{D}^{T}(\pi)_{ki}\mc{D}^{\mc{B}}(\pi)_{lj}\right)T_i\mc{B}_j.
\end{eqnarray}

Alternatively, one could obtain the p-basis by inner product decomposition as in Ref.~\cite{Li:2020gnx,Li:2020xlh}, where one first finds the irreducible Lorentz factors $\mc{B}^{[\lambda]}{}_x$ and gauge factors $T_G^{[\lambda]}{}_x$ and then combine them with the CG coefficients of the permutation group as
\begin{eqnarray}\label{eq:master}
\mc{O}_{(\lambda,x),\zeta} = \sum_{x_1,x_2} C_{(\lambda,x),\zeta}^{(\lambda_{1},x_1),(\lambda_{2},x_2)}\mathcal{B}^{\lambda_{1}}_{x_1} T^{\lambda_2}_{{\rm SU3},x_2},
\end{eqnarray}
To get the final form as $\mc{O}_{(\lambda,1),\zeta} = \mc{Y}^{[\lambda]}_1\circ\mc{O}_\zeta$, we could simply find the coordinate of $\mc{O}_\zeta$ in the p-basis which encode the combination of p-basis amplitudes with $[\lambda]$ symmetry one should get from the projection of $\mc{Y}^{[\lambda]}_1$. It is then straightforward to select a set $\{\mc{O}_\zeta\}$ with independent projections, which we refer to as the ``desymmetrization'' process in \cite{Li:2020xlh}.

When we contract the flavor tensor with the Wilson coefficient tensor $C_{prs}$, the Young symmetrizer could be interpreted as acting on $C_{prs}$ with inverse permutations
\begin{eqnarray}
\sum_{prs} C_{prs}\left( {\cal Y}^{[\lambda]}\circ{\cal O}^{prs}\right) = \sum_{prs}\left( \overline{{\cal Y}^{[\lambda]}}\circ C_{prs}\right){\cal O}^{prs},
\end{eqnarray}
Thus the result can also be interpreted as the monomial operator $\mc{O}$ with symmetrized Wilson coefficient tensor.
Note that instead of looking for flavor relations among the Wilson coefficients, we only define Wilson coefficients for the independent entries of the irreducible coefficient tensor $\overline{{\cal Y}^{[\lambda]}}\circ C_{prs}$, given by the SSYT of the Young symmetrizer. For example, the independent Wilson coefficients of a $[\lambda]=[2,1]$ symmetrized p-basis operator $\mc{Y}\left[\tiny \young(pr,s)\right]\circ\mc{O}^{prs}$ are given by the 8 SSYT's:
\eq{
    & \young(11,2)\sim C_{112}, \quad \young(11,3)\sim C_{113}, \quad \young(12,2)\sim C_{122}, \quad \young(12,3)\sim C_{123}, \\
    & \young(13,2)\sim C_{132}, \quad \young(13,3)\sim C_{133}, \quad \young(22,3)\sim C_{223}, \quad \young(23,3)\sim C_{233},
}
while all the other components in the tensor $C_{prs}$ are either 0 or related to the above ones by the tensor symmetry $[\lambda]$.

\section{Lists of Operators in LEFT}\label{sec:List}

In this section, we list all the independent dimension 5, 6, 7, 8 and 9 operators in LEFT, and the number of operators with various lepton number violation $|\Delta L|$, are summarized for each dimension in table~\ref{tab:sumdim5678} and table~\ref{tab:sumdim9}. Although the two-component Weyl spinors are used to construct the operator basis, we changed them into four-component Dirac spinors when presenting the result for user's convenience. The conversion relations between two-component Weyl spinors and four-component Dirac spinors are as follows.
\begin{align}\label{eq:conversion1}
\nu_{\rm L}=\begin{pmatrix}\nu\\0\end{pmatrix},\quad
e_{\rm L}=\left(\begin{array}{c}e\\0\end{array}\right),\quad 
e_{\rm R}=\left(\begin{array}{c}0\\e_{_\mathbb{C}}^{\dagger}\end{array}\right),\quad
u_{\rm{L}}=\begin{pmatrix}u\\0\end{pmatrix},\quad u_{\rm{R}}=\left(\begin{array}{c}0\\u_{_\mathbb{C}}^{\dagger}\end{array}\right),\quad 
d_{\rm{L}}=\begin{pmatrix}d\\0\end{pmatrix},\quad
d_{\rm R}=\left(\begin{array}{c}0\\d_{_\mathbb{C}}^{\dagger}\end{array}\right).\\
\bar{\nu}_{\rm L}=\left(0\,,\,\nu^{\dagger}\right),\quad
\bar{e}_{\rm L}=\left(0\,,\,e^{\dagger}\right),\quad 
\bar{e}_{\rm R}=\left(e_{_\mathbb{C}}\,,\,0\right),\quad
\bar{u}_{\rm{L}}=\left(0\,,\,u^{\dagger} \right),\quad \bar{u}_{\rm{R}}=\left(u_{_\mathbb{C}}\,,\,0 \right),\quad
\bar{d}_{\rm{L}}=\left(0\,,\,d^{\dagger} \right),\quad 
\bar{d}_{\rm R}=\left(d_{_\mathbb{C}}\,,\,0\right).\label{eq:conversion2}
\end{align}
In the following lists of operators, the $\psi$ ($\psi^\dagger$) in the name of each class indicates a two-component left-handed (right handed) spinor in this class, and the baryon-number and lepton-number violation pattern of each type is listed next to the type as $(\Delta B, \Delta L)$. We label the (anti)fundamental representation of group  $SU(3)_C$ and the adjoint representation of $SU(3)_C$ by the indices $\{a,b,c,d,e,f\}$ and $\{A,B,C,D\}$, respectively, and the indices for flavor of the fermion fields are denoted by $\{p,r,s,t,u,v\}$. 

After applying the method introduced in section \ref{sec:CC} to LEFT, we are able to obtain the complete basis of independent operators in LEFT at various mass dimensions. In order to compare our result with others, we change the two-component spinors in our building blocks to four-component spinors using the correspondence in eq.~\eqref{eq:conversion1} and ~\eqref{eq:conversion2}. Furthermore, since the operator basis in LEFT can be different in the literature, the equivalence relations due to Schouten Identity, Fierz Identity and IBP need to be taken into account in the comparison. 

To show the conversion between two-component spinors and four-component spinors and compare the result with others, let us take type $\bar{d}_{\rm R} d_{\rm L} \bar{u}_{\rm R} u_{\rm L}$ as an example. Operators of this type are written as two-component spinors at first,
\begin{align}
		
		& \left(d_{_\mathbb{C}}{}_{p}^{a} d_{rb}\right) \left(u_{_\mathbb{C}}{}_{s}^{b} u_{ta}\right),\label{eq:dduu1}
		
		\\& \left(d_{_\mathbb{C}}{}_{p}^{a} d_{ra}\right) \left(u_{_\mathbb{C}}{}_{s}^{c} u_{tc}\right),
		
		\\& \left(d_{rb} u_{ta}\right) \left(d_{_\mathbb{C}}{}_{p}^{a} u_{_\mathbb{C}}{}_{s}^{b}\right),
		
		\\& \left(d_{ra} u_{tc}\right) \left(d_{_\mathbb{C}}{}_{p}^{a} u_{_\mathbb{C}}{}_{s}^{c}\right).\label{eq:dduu4}
    
\end{align}
These operators can be converted to the (chiral) four-component spinor notation eq.~(\ref{eq:conversion1}, \ref{eq:conversion2}) using
\eq{\label{eq:bilinear}
	\bar{\Psi}_1\Psi_2=&\chi_1^{\alpha}\xi_{2\alpha}+\xi^{\dagger}_{1\dot{\alpha}}\chi^{\dagger\dot{\alpha}}_2\;,\\
	\Psi^{\rm{T}}_1C\Psi_2=&\xi_1^{\alpha}\xi_{2\alpha}+\chi^{\dagger}_{1\dot{\alpha}}\chi^{\dagger\dot{\alpha}}_2\;,\\
	\bar{\Psi}_1C\bar{\Psi}_2^{\rm{T}}=&\xi^{\dagger}_{1\dot{\alpha}}\xi^{\dagger\dot{\alpha}}_2+\chi^{\alpha}_1\chi_{2\alpha}\;,
}
where $\Psi$s are four-component spinors and $\xi$s, $\chi$s are left-handed two-component spinors,
\begin{align}
	\Psi=\left(\begin{array}{c} \xi_{\alpha}\\\chi^{\dagger\dot{\alpha}} \end{array} \right),\quad \bar{\Psi}=\Psi^{\dagger}\gamma^0=\left(\chi^{\alpha},\;\xi^{\dagger}_{\dot{\alpha}} \right)\;,
\end{align}
and $C=i\gamma^0\gamma^2=\begin{pmatrix} \epsilon_{\alpha\beta}&0\\0&\epsilon^{\dot{\alpha}\dot{\beta}}\end{pmatrix}=\begin{pmatrix} -\epsilon^{\alpha\beta}&0\\0&-\epsilon_{\dot{\alpha}\dot{\beta}}\end{pmatrix}$. 
Taking eq.~\eqref{eq:dduu1} as an example, the pairs of fields in each parenthesis can be translated into four-components ones of the form $\bar\Psi \Psi$ using eq.~\eqref{eq:conversion1} and \eqref{eq:conversion2}, with the first line in eq.~\eqref{eq:bilinear}, the two-component notation is directly translated into the four component one: $\left(\overline{d}_{\rm R}{}_{p}^{a} d_{\rm L}{}_{rb}\right) \left(\overline{u}_{\rm R}{}_{s}^{b} u_{\rm L}{}_{ta}\right)$. For the case in eq.~\eqref{eq:dduu4} where the pairs of fields in each parenthesis have the forms of $\Psi\Psi$ or $\bar{\Psi}\bar\Psi$ after conversion to four-component ones, the charge conjugate operators $C$ and the proper transposition of spinors need to be inserted properly following the last two lines in eq.~\eqref{eq:bilinear} yielding $\left(d_{\rm L}^{\rm T}{}_{ra} C u_{\rm L}{}_{tc}\right) \left(\overline{d}_{\rm R}{}_{p}^{a} C \overline{u}_{\rm R}^{\rm T}{}_{s}^{c}\right)$.
More examples of such conversions can be found in the appendix of Ref.~\cite{Li:2020gnx}. Thus the operators of type $\bar{d}_{\rm R} d_{\rm L} \bar{u}_{\rm R} u_{\rm L}$ can be presented in four-component spinors as
\begin{align}
     \left(\overline{d}_{\rm R}{}_{p}^{a} d_{\rm L}{}_{rb}\right) \left(\overline{u}_{\rm R}{}_{s}^{b} u_{\rm L}{}_{ta}\right) &= \left(d_{_\mathbb{C}}{}_{p}^{a} d_{rb}\right) \left(u_{_\mathbb{C}}{}_{s}^{b} u_{ta}\right),
		
		\\ \left(\overline{d}_{\rm R}{}_{p}^{a} d_{\rm L}{}_{ra}\right) \left(\overline{u}_{\rm R}{}_{s}^{c} u_{\rm L}{}_{tc}\right) &= \left(d_{_\mathbb{C}}{}_{p}^{a} d_{ra}\right) \left(u_{_\mathbb{C}}{}_{s}^{c} u_{tc}\right),
		
		\\ \left(d_{\rm L}^{\rm T}{}_{rb} C u_{\rm L}{}_{ta}\right) \left(\overline{d}_{\rm R}{}_{p}^{a} C \overline{u}_{\rm R}^{\rm T}{}_{s}^{b}\right) &= \left(d_{rb} u_{ta}\right) \left(d_{_\mathbb{C}}{}_{p}^{a} u_{_\mathbb{C}}{}_{s}^{b}\right),
		
		\\ \left(d_{\rm L}^{\rm T}{}_{ra} C u_{\rm L}{}_{tc}\right) \left(\overline{d}_{\rm R}{}_{p}^{a} C \overline{u}_{\rm R}^{\rm T}{}_{s}^{c}\right) &= \left(d_{ra} u_{tc}\right) \left(d_{_\mathbb{C}}{}_{p}^{a} u_{_\mathbb{C}}{}_{s}^{c}\right).
\end{align}
These operators on the left-hand side are the final form presented in our result in eq.~(\ref{ty:bdRdLbuRuL}), where the transpose symbol $\rm{T}$ is omitted for simplicity. 

To compare our bases with their counterparts in Ref.~\cite{Jenkins:2017jig}, Fierz identities in the following equations are needed:
\begin{eqnarray}
    \left(d_{\rm L}^{\rm T} C u_{\rm L}\right) \left(\overline{d}_{\rm R} C \overline{u}_{\rm R}^{\rm T}\right) &=& -\left(\overline{d}_{\rm R} u_{\rm L}\right) \left(\overline{u}_{\rm R} d_{\rm L}\right)-\left(\overline{d}_{\rm R} d_{\rm L}\right) \left(\overline{u}_{\rm R} u_{\rm L}\right), \\
    \sum_{A} (T^{A})_{ab}(T^{A})_{cd} &=& \delta_{ad}\delta_{cb}-\frac{1}{N}\delta_{ab}\delta_{cd},
\end{eqnarray}
and it can be shown that the following basis operators are equivalent to ours 
\begin{eqnarray}
    &&\left(\overline{d}_{\rm R}{}_{p} d_{\rm L}{}_{r}\right) \left(\overline{u}_{\rm R}{}_{s} u_{\rm L}{}_{t}\right), \\
    &&\left(\overline{d}_{\rm R}{}_{p} T^A d_{\rm L}{}_{r}\right) \left(\overline{u}_{\rm R}{}_{s} T^A u_{\rm L}{}_{t}\right), \\
    &&\left(\overline{d}_{\rm R}{}_{p} u_{\rm L}{}_{t}\right) \left(\overline{u}_{\rm R}{}_{s} d_{\rm L}{}_{r}\right), \\
    &&\left(\overline{d}_{\rm R}{}_{p} T^A u_{\rm L}{}_{t}\right) \left(\overline{u}_{\rm R}{}_{s} T^A d_{\rm L}{}_{r}\right),
\end{eqnarray}
which are exactly the Hermitian conjugate of operators $\mathcal{O}^{S1,RR}_{ud}$, $\mathcal{O}^{S8,RR}_{ud}$, $\mathcal{O}^{S1,RR}_{uddu}$ and $\mathcal{O}^{S8,RR}_{uddu}$ in Ref.~\cite{Jenkins:2017jig}.

In addition to the conversion between two-component and four-component spinors, we used some identities of $\sigma$ matrices to convert spinor indices to Lorentz indices, for example, 
\begin{eqnarray}\label{eq:sigmaeg}
    F_L{}^{\alpha\beta} F_L{}_{\alpha\beta} = -\dfrac{1}{4} F_L{}_{\mu\nu} F_L{}_{\rho\lambda} \left(\sigma^{\mu\nu}\right)^{\alpha\beta} \left(\sigma^{\rho\lambda}\right)_{\alpha\beta} = \dfrac{1}{4} F_L{}_{\mu\nu} F_L{}_{\rho\lambda} \rm{Tr} \left(\sigma^{\mu} \bar{\sigma}^{\nu} \sigma^{\rho} \bar{\sigma}^{\lambda}\right) 
\end{eqnarray}
where 
\begin{eqnarray}
    F_{{\rm L}\alpha\beta} = \frac{i}{2}F_{\mu\nu}\left(\sigma^{\mu\nu}\right)_{\alpha\beta} = \frac{i}{2}F_L{}_{\mu\nu}\left(\sigma^{\mu\nu}\right)_{\alpha\beta},
\end{eqnarray}
and the following identities are used,
\begin{eqnarray}
	\left(\sigma^{\mu}\bar{\sigma}^{\nu}\right)_{\alpha}{}^{\beta}=&g^{\mu\nu}\delta^{\beta}_{\alpha}-i\left(\sigma^{\mu\nu}\right)_{\alpha}{}^{\beta},\label{eq:2si}\\
	\left(\bar{\sigma}^{\mu}\sigma^{\nu}\right)^{\dot{\alpha}}{}_{\dot{\beta}}=&g^{\mu\nu}\delta^{\dot{\alpha}}_{\dot{\beta}}-i\left(\bar{\sigma}^{\mu\nu}\right)^{\dot{\alpha}}{}_{\dot{\beta}}.\label{eq:2sibar}
\end{eqnarray}
The trace of $\sigma$ matrices in eq.~(\ref{eq:sigmaeg}) can be simplified in a systematic way using our code and yields
\begin{eqnarray}
    {\rm Tr} \left(\sigma^{\mu}\bar{\sigma}^{\nu}\sigma^{\rho}\bar{\sigma}^{\lambda}\right)=2g^{\mu\nu}g^{\rho\lambda}-2g^{\mu\rho}g^{\nu\lambda}+2g^{\nu\rho}g^{\mu\lambda}+2 i \epsilon^{\mu\nu\rho\lambda}.
\end{eqnarray}
Finally we get
\begin{eqnarray}
    F_L{}^{\alpha\beta} F_L{}_{\alpha\beta} = \dfrac{1}{2} F_L{}^{\mu}{}_{\mu} F_L{}^{\rho}{}_{\rho} - \dfrac{1}{2} F_L{}^{\mu\nu} F_L{}_{\mu\nu} + \dfrac{1}{2} F_L{}^{\mu\nu} F_L{}_{\nu\mu} - F_L{}^{\mu\nu} F_L{}_{\mu\nu} = -2 F_L{}^{\mu\nu} F_L{}_{\mu\nu}.
\end{eqnarray}

\begin{table}[t]
	\begin{center}
		\begin{tabular}{|c|cc|cc|ccc|}
			\hline
			\text{Fields} & $SU(2)_{l}\times SU(2)_{r}$	& $h$ & $SU(3)_{C}$ & $U(1)_{\rm EM}$ &  Flavor & $B$ & $L$ \tabularnewline
			\hline
			$G_{\rm L\alpha\beta}^A$   & $\left(1,0\right)$  & $-1$    & $\boldsymbol{8}$ & 0  & $1$ & 0 & 0 \tabularnewline
			$F_{\rm L\alpha\beta}$   & $\left(1,0\right)$    & $-1$        & $\boldsymbol{1}$ & 0  & $1$ & 0 & 0 \tabularnewline
			\hline
			$\nu_{\alpha}$     & $\left(\frac{1}{2},0\right)$  & $-\frac12$  & $\boldsymbol{1}$ & $0$  & $n_e$ & 0 & $1$ \tabularnewline
			$e_{\alpha}$ & $\left(\frac{1}{2},0\right)$ & $-\frac12$   & $\boldsymbol{1}$ & $-1$  & $n_e$ & 0 & $1$ \tabularnewline
			$e_{_\mathbb{C}\alpha}$ & $\left(\frac{1}{2},0\right)$ & $-\frac12$   & $\boldsymbol{1}$ & $1$  & $n_e$ & 0 & $-1$ \tabularnewline
			$u_{\alpha a}$     & $\left(\frac{1}{2},0\right)$ & $-\frac12$   & $\boldsymbol{3}$ & $\frac23$  & $n_u$ & $\frac13$ & 0 \tabularnewline
			$u_{_\mathbb{C}\alpha}^a$ & $\left(\frac{1}{2},0\right)$ & $-\frac12$   & $\overline{\boldsymbol{3}}$ & $-\frac23$  & $n_u$ & $-\frac13$ & 0 \tabularnewline
			$d_{\alpha a}$     & $\left(\frac{1}{2},0\right)$ & $-\frac12$   & $\boldsymbol{3}$ & $-\frac13$  & $n_d$ & $\frac13$ & 0 \tabularnewline
			$d_{_\mathbb{C}\alpha}^a$ & $\left(\frac{1}{2},0\right)$ & $-\frac12$   & $\overline{\boldsymbol{3}}$ & $\frac13$  & $n_d$ & $-\frac13$ & $0$ \tabularnewline
			\hline
		\end{tabular}
		\caption{\label{tab:LEFT-field-content}
			The field content of the LEFT, along with their representations under the Lorentz and gauge symmetries. The representation under Lorentz group is denoted by $(j_l,j_r)$, while the helicity of the field is given by $h = j_r-j_l$ .
			The numbers of lepton flavors, u-type quark flavors and d-type quark flavors are denoted as $n_e$, $n_u$ and $n_d$ respectively with $n_e=3$, $n_u=2$ and $n_d=3$ in LEFT. Their global charges, baryon number $B$ and lepton number $L$ are also listed. All of the fields are accompanied with their Hermitian conjugates that are omitted, $(F_{\rm L \alpha\beta})^\dagger = F_{\rm R \dot\alpha\dot\beta}$ for gauge bosons and $(\psi_\alpha)^\dagger = (\psi^\dagger)_{\dot\alpha}$ for fermions, which are under the conjugate representations of all the groups. }
	\end{center}
\end{table}

\begin{table}
	\begin{align*}
		\begin{array}{cc|c|c|c|c|c}
			\hline
			\multicolumn{7}{c}{\text{Dim-5 operators}}\\
			\hline
			N & (n,\tilde{n}) & \text{Classes} & \mathcal{N}_{\text{type}} & \mathcal{N}_{\text{term}} & \mathcal{N}_{\text{operator}} & \text{Equations}\\
			\hline
			3 & (2,0) & F_{\rm{L}} \psi^2    +h.c. & 10+0+2+0 & 12 & 76 & \multirow{1}*{(\ref{cl:FLpsiL2f}-\ref{cl:FLpsiL2l})} \\
			\hline\hline
			\multicolumn{7}{c}{\text{Dim-6 operators}}\\
			\hline
			N & (n,\tilde{n}) & \text{Classes} & \mathcal{N}_{\text{type}} & \mathcal{N}_{\text{term}} & \mathcal{N}_{\text{operator}} & \text{Equations}\\
			\hline
			3 & (3,0) & F_{\rm{L}}^3    +h.c. & 2+0+0+0 & 2 & 2 & (\ref{cl:FL3f}) \\
			\hline
			4 & (2,0) & \psi^4    +h.c. & 14+12+8+2 & 78 & 2362 & \multirow{1}*{(\ref{cl:psiL4f}-\ref{cl:psiL4l})} \\
			\cline{2-7}
			& (1,1) & \psi^2 \psi^\dagger{}^2 & 40+20+12+0 & 84 & 3511 & \multirow{1}*{(\ref{cl:psiL2psiR2f}-\ref{cl:psiL2psiR2l})} \\
			\hline
			\multicolumn{2}{c|}{\text{Total}} & 5 & 56+32+20+2 & 164 & 5875 & \\
			\hline\hline
			\multicolumn{7}{c}{\text{Dim-7 operators}}\\
			\hline
			N & (n,\tilde{n}) & \text{Classes} & \mathcal{N}_{\text{type}} & \mathcal{N}_{\text{term}} & \mathcal{N}_{\text{operator}} & \text{Equations}\\
			\hline
			4 & (3,0) & F_{\rm{L}}^2 \psi^2    +h.c. & 16+0+4+0 & 32 & 216 & \multirow{1}*{(\ref{cl:FL2psiL2f}-\ref{cl:FL2psiL2l})} \\
			\cline{2-7}
			& (2,1) & F_{\rm{L}}^2 \psi^\dagger{}^2    +h.c. & 16+0+4+0 & 24 & 164 & \multirow{1}*{(\ref{cl:FL2psiR2f}-\ref{cl:FL2psiR2l})} \\
			& & \psi^3 \psi^\dagger{} D   +h.c. & 50+32+22+0 & 120 & 4738 & \multirow{1}*{(\ref{cl:psiL3psiRDf}-\ref{cl:psiL3psiRDl})} \\
			\hline
			\multicolumn{2}{c|}{\text{Total}} & 6 & 82+32+30+0 & 176 & 5118 & \\
			\hline\hline
			\multicolumn{7}{c}{\text{Dim-8 operators}}\\
			\hline
			N & (n,\tilde{n}) & \text{Classes} & \mathcal{N}_{\text{type}} & \mathcal{N}_{\text{term}} & \mathcal{N}_{\text{operator}} & \text{Equations}\\
			\hline
			4 & (4,0) & F_{\rm{L}}^4    +h.c. & 8+0+0+0 & 14 & 14 & \multirow{1}*{(\ref{cl:FL4f}-\ref{cl:FL4l})} \\
			\cline{2-7}
			& (3,1) & F_{\rm{L}}^2 \psi \psi^\dagger{} D   +h.c. & 16+0+0+0 & 16 & 104 & \multirow{1}*{(\ref{cl:FL2psiLpsiRDf}-\ref{cl:FL2psiLpsiRDl})} \\
			&  & \psi^4 D^2   +h.c. & 14+12+8+2 & 120 & 3552 & \multirow{1}*{(\ref{cl:psiL4D2f}-\ref{cl:psiL4D2l})} \\
			\cline{2-7}
			& (2,2) & F_{\rm{L}}^2 F_{\rm{R}}^2 & 7+0+0+0 & 9 & 9 & \multirow{1}*{(\ref{cl:FL2FR2f}-\ref{cl:FL2FR2l})} \\
			& & F_{\rm{L}} F_{\rm{R}} \psi \psi^\dagger{} D & 22+0+0+0 & 30 & 210 & \multirow{1}*{(\ref{cl:FLFRpsiLpsiRDf}-\ref{cl:FLFRpsiLpsiRDl})} \\
			& & \psi^2 \psi^\dagger{}^2   D^2 & 40+20+12+0 & 168 & 7061 & \multirow{1}*{(\ref{cl:psiL2psiR2D2f}-\ref{cl:psiL2psiR2D2l})} \\
			\hline
			5 & (3,0) & F_{\rm{L}} \psi^4   +h.c. & 26+24+14+2 & 290 & 8024 & \multirow{1}*{(\ref{cl:FLpsiL4f}-\ref{cl:FLpsiL4l})} \\
			\cline{2-7}
			& (2,1) & F_{\rm{L}} \psi^2 \psi^\dagger{}{}^2   +h.c. & 148+80+44+0 & 408 & 16084 & \multirow{1}*{(\ref{cl:FLpsiL2psiR2f}-\ref{cl:FLpsiL2psiR2l})} \\
			\hline
			\multicolumn{2}{c|}{\text{Total}} & 13 & 281+136+78+4 & 1055 & 35058 & \\
			\hline
		\end{array}
    \end{align*}
	\caption{The complete statistics of dimension 5, 6, 7, 8 LEFT operators.
		$N$ in the leftmost column shows the number of particles. $(n,\tilde{n})$ are the numbers of $\epsilon$
		and $\tilde{\epsilon}$ in the Lorentz structure. $\mathcal{N}_{\rm type}$, $\mathcal{N}_{\rm term}$, and $\mathcal{N}_{\rm operator}$
		show the number of types, terms and Hermitian operators respectively (independent conjugates
		are counted), while the numbers under $\mathcal{N}_{\rm type}$ describe the sum of each possible $|\Delta L|$
		types/operators with $\mathcal{N}=\mathcal{N}(|\Delta L|=0)+\mathcal{N}(|\Delta L|=1)+\mathcal{N}(|\Delta L|=2)+\mathcal{N}(|\Delta L|=4)$. The links in the rightmost column refer to the list(s) of the terms in given
		classes.
	}
	\label{tab:sumdim5678}
\end{table}
\begin{table}
	\begin{align*}
\begin{array}{cc|c|c|c|c|c}
			\hline
			\multicolumn{7}{c}{\text{Dim-9 operators}}\\
			\hline
			N & (n,\tilde{n}) & \text{Classes} & \mathcal{N}_{\text{type}} & \mathcal{N}_{\text{term}} & \mathcal{N}_{\text{operator}} & \text{Equations}\\
			\hline
			4 & (3,2) & F_{\rm{L}}{}^2 \psi^2   D^2    +h.c. & 16+0+4+0+0+0 & 56 & 380 & \multirow{1}*{(\ref{cl:FL2psiL2D2f}-\ref{cl:FL2psiL2D2l})} \\
			& & F_{\rm{L}} F_{\rm{R}} \psi^2   D^2    +h.c. & 20+0+4+0+0+0 & 32 & 216 & \multirow{1}*{(\ref{cl:FLFRpsiL2D2f}-\ref{cl:FLFRpsiL2D2l})} \\
			& & F_{\rm{L}}{}^2 \psi^\dagger{}{}^2   D^2    +h.c. & 16+0+4+0+0+0 & 32 & 216 & \multirow{1}*{(\ref{cl:FL2psiR2D2f}-\ref{cl:FL2psiR2D2l})} \\
			& & \psi^3 \psi^\dagger{}   D^3    +h.c. & 50+32+22+0+0+0 & 234 & 9344 & \multirow{1}*{(\ref{cl:psiL3psiRD3f}-\ref{cl:psiL3psiRD3l})} \\
			\hline
			5 & (4,0) & F_{\rm{L}}{}^3 \psi^2    +h.c. & 22+0+6+0+0+0 & 84 & 542 & \multirow{1}*{(\ref{cl:FL3psiL2f}-\ref{cl:FL3psiL2l})} \\
			\cline{2-7}
			& (3,1) & F_{\rm{L}}{}^3 \psi^\dagger{}^2    +h.c. & 10+0+2+0+0+0 & 16 & 108 & \multirow{1}*{(\ref{cl:FL3psiR2f}-\ref{cl:FL3psiR2l})} \\
			& & F_{\rm{L}}{} \psi^3 \psi^\dagger D   +h.c. & 94+64+38+0+0+0 & 1150 & 43002 & \multirow{1}*{(\ref{cl:FLpsiL3psiRDf}-\ref{cl:FLpsiL3psiRDl})} \\
			\cline{2-7}
			& (2,2) & F_{\rm{L}} F_{\rm{R}}{}^2 \psi^2    +h.c. & 32+0+8+0+0+0 & 68 & 434 & \multirow{1}*{(\ref{cl:FLFR2psiL2f}-\ref{cl:FLFR2psiL2l})} \\
			& & F_{\rm{R}} \psi^3 \psi^\dagger   D    +h.c. & 94+64+38+0+0+0 & 862 & 32208 & \multirow{1}*{(\ref{cl:FRpsiL3psiRDf}-\ref{cl:FRpsiL3psiRDl})} \\
			\hline
			6 & (3,0) & \psi^6    +h.c. & 30+42+24+10+8+2 & 908 & 88136 & \multirow{1}*{(\ref{cl:psiL6f}-\ref{cl:psiL6l})} \\
			\cline{2-7}
			& (2,1) & \psi^4 \psi^\dagger{}^2    +h.c. & 272+318+184+34+12+0 & 3012 & 529998 & \multirow{1}*{(\ref{cl:psiL4psiR2f}-\ref{cl:psiL4psiR2l})} \\
			\hline
			\multicolumn{2}{c|}{\text{Total}} & 22 & 656+520+334+44+20+2 & 6454 & 704584 & \\
			\hline
		\end{array}
	\end{align*}
	\caption{The complete statistics of dimension 9 LEFT operators. The numbers under $\mathcal{N}_{\rm type}$ describe the sum of each possible $|\Delta L|$ types/operators with $\mathcal{N}=\mathcal{N}(|\Delta L|=0)+\mathcal{N}(|\Delta L|=1)+\mathcal{N}(|\Delta L|=2)+\mathcal{N}(|\Delta L|=3)+\mathcal{N}(|\Delta L|=4)+\mathcal{N}(|\Delta L|=6)$.
	}
	\label{tab:sumdim9}
\end{table}

\subsection{Lists of the Dim-5 Operators}

\indent \indent \underline{Class $F_{\rm{L}} \psi {}^2    $}: 6 types

\begin{align}

&\begin{array}{c|l}\label{cl:FLpsiL2f}

\mathcal{O}_{G_{\rm{L}} \bar{u}_{\rm R} u_{\rm L}    }(0,0)

& i \left(\lambda ^A\right)_a^b G_{\rm{L}}^{A}{}^{\mu }{}^{\nu } \left(\overline{u}_{\rm R}{}_{p}^{a} \sigma _{\mu }{}_{\nu } u_{\rm L}{}_{rb}\right)

\end{array}\\

&\begin{array}{c|l}

\mathcal{O}_{G_{\rm{L}} \bar{d}_{\rm R} d_{\rm L}    }(0,0)

& i \left(\lambda ^A\right)_a^b G_{\rm{L}}^{A}{}^{\mu }{}^{\nu } \left(\overline{d}_{\rm R}{}_{p}^{a} \sigma _{\mu }{}_{\nu } d_{\rm L}{}_{rb}\right)

\end{array}\\

&\begin{array}{c|l}

\mathcal{O}_{F_{\rm{L}} \bar{u}_{\rm R} u_{\rm L}    }(0,0)

& i F_{\rm{L}}{}^{\mu }{}^{\nu } \left(\overline{u}_{\rm R}{}_{p}^{a} \sigma _{\mu }{}_{\nu } u_{\rm L}{}_{ra}\right)

\end{array}\\

&\begin{array}{c|l}

\mathcal{O}_{F_{\rm{L}} \bar{d}_{\rm R} d_{\rm L}    }(0,0)

& i F_{\rm{L}}{}^{\mu }{}^{\nu } \left(\overline{d}_{\rm R}{}_{p}^{a} \sigma _{\mu }{}_{\nu } d_{\rm L}{}_{ra}\right)

\end{array}\\

&\begin{array}{c|l}

\mathcal{O}_{F_{\rm{L}} \bar{e}_{\rm R} e_{\rm L}    }(0,0)

& i F_{\rm{L}}{}^{\mu }{}^{\nu } \left(\overline{e}_{\rm R}{}_{p} \sigma _{\mu }{}_{\nu } e_{\rm L}{}_{r}\right)

\end{array}\\

&\begin{array}{c|l}\label{cl:FLpsiL2l}

\mathcal{O}_{F_{\rm{L}} \nu _{\rm L}^2    }(0,2)

&\mathcal{Y}\left[\tiny{\young(p,r)}\right]i F_{\rm{L}}{}^{\mu }{}^{\nu } \left(\nu _{\rm L}{}_{p} C \sigma _{\mu }{}_{\nu } \nu _{\rm L}{}_{r}\right)

\end{array}

\end{align}

\subsection{Lists of the Dim-6 Operators}

\indent \indent \underline{Class $F_{\rm{L}}{}^3      $}: 1 types

\begin{align}
	
	&\begin{array}{c|l}\label{cl:FL3f}
		
		\mathcal{O}_{G_{\rm{L}}{}^3      }(0,0)
		
		& f^{ABC} G_{\rm{L}}^{A}{}_{\mu }{}_{\nu } G_{\rm{L}}^{B}{}^{\mu }{}{}_{\lambda } G_{\rm{L}}^{C}{}^{\nu }{}^{\lambda }
		
	\end{array}
	
\end{align}

\underline{Class $  \psi {}^4    $}: 18 types

\begin{align}
	
	&\begin{array}{c|l}\label{cl:psiL4f}
		
		\multirow{4}*{$\mathcal{O}_{  \bar{u}_{\rm R}^2 u_{\rm L}^2    }^{(1\sim 4)}(0,0)$}
		
		&\mathcal{Y}\left[\tiny{\young(p,r)},\tiny{\young(s,t)}\right]\left(u_{\rm L}{}_{sa} C u_{\rm L}{}_{tb}\right) \left(\overline{u}_{\rm R}{}_{p}^{a} C \overline{u}_{\rm R}{}_{r}^{b}\right)
		
		\\&\mathcal{Y}\left[\tiny{\young(p,r)},\tiny{\young(s,t)}\right]\left(\overline{u}_{\rm R}{}_{p}^{a} u_{\rm L}{}_{sa}\right) \left(\overline{u}_{\rm R}{}_{r}^{b} u_{\rm L}{}_{tb}\right)
		
		\\&\mathcal{Y}\left[\tiny{\young(pr)},\tiny{\young(st)}\right]\left(u_{\rm L}{}_{sa} C u_{\rm L}{}_{tb}\right) \left(\overline{u}_{\rm R}{}_{p}^{a} C \overline{u}_{\rm R}{}_{r}^{b}\right)
		
		\\&\mathcal{Y}\left[\tiny{\young(pr)},\tiny{\young(st)}\right]\left(\overline{u}_{\rm R}{}_{p}^{a} u_{\rm L}{}_{sa}\right) \left(\overline{u}_{\rm R}{}_{r}^{b} u_{\rm L}{}_{tb}\right)
		
	\end{array}\\
	
	&\begin{array}{c|l}
		
		\multirow{2}*{$\mathcal{O}_{  d_{\rm L} e_{\rm L} u_{\rm L}^2    }^{(1,2)}(1,1)$}
		
		&\mathcal{Y}\left[\tiny{\young(st)}\right]\epsilon ^{abc} \left(e_{\rm L}{}_{r} C u_{\rm L}{}_{tc}\right) \left(d_{\rm L}{}_{pa} C u_{\rm L}{}_{sb}\right)
		
		\\&\mathcal{Y}\left[\tiny{\young(s,t)}\right]\epsilon ^{abc} \left(d_{\rm L}{}_{pa} C e_{\rm L}{}_{r}\right) \left(u_{\rm L}{}_{sb} C u_{\rm L}{}_{tc}\right)
		
	\end{array}\\
	
	&\begin{array}{c|l}
		
		\multirow{4}*{$\mathcal{O}_{  \bar{d}_{\rm R} d_{\rm L} \bar{u}_{\rm R} u_{\rm L}    }^{(1\sim 4)}(0,0)$}
		
		& \left(\overline{d}_{\rm R}{}_{p}^{a} d_{\rm L}{}_{rb}\right) \left(\overline{u}_{\rm R}{}_{s}^{b} u_{\rm L}{}_{ta}\right)
		
		\\& \left(\overline{d}_{\rm R}{}_{p}^{a} d_{\rm L}{}_{ra}\right) \left(\overline{u}_{\rm R}{}_{s}^{c} u_{\rm L}{}_{tc}\right)
		
		\\& \left(d_{\rm L}{}_{rb} C u_{\rm L}{}_{ta}\right) \left(\overline{d}_{\rm R}{}_{p}^{a} C \overline{u}_{\rm R}{}_{s}^{b}\right)
		
		\\& \left(d_{\rm L}{}_{ra} C u_{\rm L}{}_{tc}\right) \left(\overline{d}_{\rm R}{}_{p}^{a} C \overline{u}_{\rm R}{}_{s}^{c}\right)
		
	\end{array}\\
	
	&\begin{array}{c|l}
		
		\multirow{2}*{$\mathcal{O}_{  \bar{e}_{\rm R} e_{\rm L} \bar{u}_{\rm R} u_{\rm L}    }^{(1,2)}(0,0)$}
		
		& \left(\overline{e}_{\rm R}{}_{p} e_{\rm L}{}_{r}\right) \left(\overline{u}_{\rm R}{}_{s}^{a} u_{\rm L}{}_{ta}\right)
		
		\\& \left(e_{\rm L}{}_{r} C u_{\rm L}{}_{ta}\right) \left(\overline{e}_{\rm R}{}_{p} C \overline{u}_{\rm R}{}_{s}^{a}\right)
		
	\end{array}\\
	
	&\begin{array}{c|l}
		
		\multirow{2}*{$\mathcal{O}_{  \nu _{\rm L}^2 \bar{u}_{\rm R} u_{\rm L}    }^{(1,2)}(0,2)$}
		
		&\mathcal{Y}\left[\tiny{\young(s,t)}\right]\left(u_{\rm L}{}_{ra} C \nu _{\rm L}{}_{t}\right) \left(\overline{u}_{\rm R}{}_{p}^{a} \nu _{\rm L}{}_{s}\right)
		
		\\&\mathcal{Y}\left[\tiny{\young(st)}\right]\left(\nu _{\rm L}{}_{s} C \nu _{\rm L}{}_{t}\right) \left(\overline{u}_{\rm R}{}_{p}^{a} u_{\rm L}{}_{ra}\right)
		
	\end{array}\\
	
	&\begin{array}{c|l}
		
		\multirow{2}*{$\mathcal{O}_{  d_{\rm L}^2 \nu _{\rm L} u_{\rm L}    }^{(1,2)}(1,1)$}
		
		&\mathcal{Y}\left[\tiny{\young(pr)}\right]\epsilon ^{abc} \left(d_{\rm L}{}_{rb} C \nu _{\rm L}{}_{t}\right) \left(d_{\rm L}{}_{pa} C u_{\rm L}{}_{sc}\right)
		
		\\&\mathcal{Y}\left[\tiny{\young(p,r)}\right]\epsilon ^{abc} \left(d_{\rm L}{}_{pa} C d_{\rm L}{}_{rb}\right) \left(u_{\rm L}{}_{sc} C \nu _{\rm L}{}_{t}\right)
		
	\end{array}\\
	
	&\begin{array}{c|l}
		
		\multirow{2}*{$\mathcal{O}_{  \bar{d}_{\rm R} e_{\rm L} \nu _{\rm L} u_{\rm L}    }^{(1,2)}(0,2)$}
		
		& \left(u_{\rm L}{}_{sa} C \nu _{\rm L}{}_{t}\right) \left(\overline{d}_{\rm R}{}_{p}^{a} e_{\rm L}{}_{r}\right)
		
		\\& \left(e_{\rm L}{}_{r} C \nu _{\rm L}{}_{t}\right) \left(\overline{d}_{\rm R}{}_{p}^{a} u_{\rm L}{}_{sa}\right)
		
	\end{array}\\
	
	&\begin{array}{c|l}
		
		\multirow{2}*{$\mathcal{O}_{  \bar{d}_{\rm R} \bar{e}_{\rm R} \bar{u}_{\rm R}^2    }^{(1,2)}(-1,-1)$}
		
		&\mathcal{Y}\left[\tiny{\young(st)}\right]\epsilon _{abc} \left(\overline{e}_{\rm R}{}_{r} C \overline{u}_{\rm R}{}_{t}^{c}\right) \left(\overline{d}_{\rm R}{}_{p}^{a} C \overline{u}_{\rm R}{}_{s}^{b}\right)
		
		\\&\mathcal{Y}\left[\tiny{\young(s,t)}\right]\epsilon _{abc} \left(\overline{d}_{\rm R}{}_{p}^{a} C \overline{e}_{\rm R}{}_{r}\right) \left(\overline{u}_{\rm R}{}_{s}^{b} C \overline{u}_{\rm R}{}_{t}^{c}\right)
		
	\end{array}\\
	
	&\begin{array}{c|l}
		
		\multirow{2}*{$\mathcal{O}_{  d_{\rm L} \bar{e}_{\rm R} \nu _{\rm L} \bar{u}_{\rm R}    }^{(1,2)}(0,0)$}
		
		& \left(\overline{e}_{\rm R}{}_{r} d_{\rm L}{}_{pa}\right) \left(\overline{u}_{\rm R}{}_{s}^{a} \nu _{\rm L}{}_{t}\right)
		
		\\& \left(\overline{e}_{\rm R}{}_{r} \nu _{\rm L}{}_{t}\right) \left(\overline{u}_{\rm R}{}_{s}^{a} d_{\rm L}{}_{pa}\right)
		
	\end{array}\\
	
	&\begin{array}{c|l}
		
		\multirow{2}*{$\mathcal{O}_{  \bar{d}_{\rm R}^2 \nu _{\rm L} \bar{u}_{\rm R}    }^{(1,2)}(-1,1)$}
		
		&\mathcal{Y}\left[\tiny{\young(pr)}\right]\epsilon _{abc} \left(\overline{d}_{\rm R}{}_{r}^{b} \nu _{\rm L}{}_{t}\right) \left(\overline{d}_{\rm R}{}_{p}^{a} C \overline{u}_{\rm R}{}_{s}^{c}\right)
		
		\\&\mathcal{Y}\left[\tiny{\young(p,r)}\right]\epsilon _{abc} \left(\overline{u}_{\rm R}{}_{s}^{c} \nu _{\rm L}{}_{t}\right) \left(\overline{d}_{\rm R}{}_{p}^{a} C \overline{d}_{\rm R}{}_{r}^{b}\right)
		
	\end{array}\\
	
	&\begin{array}{c|l}
		
		\mathcal{O}_{  d_{\rm L}^3 \bar{e}_{\rm R}    }(1,-1)
		
		&\mathcal{Y}\left[\tiny{\young(pr,s)}\right]\epsilon ^{abc} \left(\overline{e}_{\rm R}{}_{t} d_{\rm L}{}_{sc}\right) \left(d_{\rm L}{}_{pa} C d_{\rm L}{}_{rb}\right)
		
	\end{array}\\
	
	&\begin{array}{c|l}
		
		\multirow{4}*{$\mathcal{O}_{  \bar{d}_{\rm R}^2 d_{\rm L}^2    }^{(1\sim 4)}(0,0)$}
		
		&\mathcal{Y}\left[\tiny{\young(p,r)},\tiny{\young(s,t)}\right]\left(d_{\rm L}{}_{sa} C d_{\rm L}{}_{tb}\right) \left(\overline{d}_{\rm R}{}_{p}^{a} C \overline{d}_{\rm R}{}_{r}^{b}\right)
		
		\\&\mathcal{Y}\left[\tiny{\young(p,r)},\tiny{\young(s,t)}\right]\left(\overline{d}_{\rm R}{}_{p}^{a} d_{\rm L}{}_{sa}\right) \left(\overline{d}_{\rm R}{}_{r}^{b} d_{\rm L}{}_{tb}\right)
		
		\\&\mathcal{Y}\left[\tiny{\young(pr)},\tiny{\young(st)}\right]\left(d_{\rm L}{}_{sa} C d_{\rm L}{}_{tb}\right) \left(\overline{d}_{\rm R}{}_{p}^{a} C \overline{d}_{\rm R}{}_{r}^{b}\right)
		
		\\&\mathcal{Y}\left[\tiny{\young(pr)},\tiny{\young(st)}\right]\left(\overline{d}_{\rm R}{}_{p}^{a} d_{\rm L}{}_{sa}\right) \left(\overline{d}_{\rm R}{}_{r}^{b} d_{\rm L}{}_{tb}\right)
		
	\end{array}\\
	
	&\begin{array}{c|l}
		
		\multirow{2}*{$\mathcal{O}_{  \bar{d}_{\rm R} d_{\rm L} \bar{e}_{\rm R} e_{\rm L}    }^{(1,2)}(0,0)$}
		
		& \left(\overline{e}_{\rm R}{}_{s} e_{\rm L}{}_{t}\right) \left(\overline{d}_{\rm R}{}_{p}^{a} d_{\rm L}{}_{ra}\right)
		
		\\& \left(d_{\rm L}{}_{ra} C e_{\rm L}{}_{t}\right) \left(\overline{d}_{\rm R}{}_{p}^{a} C \overline{e}_{\rm R}{}_{s}\right)
		
	\end{array}\\
	
	&\begin{array}{c|l}
		
		\multirow{2}*{$\mathcal{O}_{  \bar{d}_{\rm R} d_{\rm L} \nu _{\rm L}^2    }^{(1,2)}(0,2)$}
		
		&\mathcal{Y}\left[\tiny{\young(s,t)}\right]\left(d_{\rm L}{}_{ra} C \nu _{\rm L}{}_{t}\right) \left(\overline{d}_{\rm R}{}_{p}^{a} \nu _{\rm L}{}_{s}\right)
		
		\\&\mathcal{Y}\left[\tiny{\young(st)}\right]\left(\nu _{\rm L}{}_{s} C \nu _{\rm L}{}_{t}\right) \left(\overline{d}_{\rm R}{}_{p}^{a} d_{\rm L}{}_{ra}\right)
		
	\end{array}\\
	
	&\begin{array}{c|l}
		
		\mathcal{O}_{  \bar{d}_{\rm R}^3 e_{\rm L}    }(-1,1)
		
		&\mathcal{Y}\left[\tiny{\young(pr,s)}\right]\epsilon _{abc} \left(\overline{d}_{\rm R}{}_{s}^{c} e_{\rm L}{}_{t}\right) \left(\overline{d}_{\rm R}{}_{p}^{a} C \overline{d}_{\rm R}{}_{r}^{b}\right)
		
	\end{array}\\
	
	&\begin{array}{c|l}
		
		\multirow{2}*{$\mathcal{O}_{  \bar{e}_{\rm R}^2 e_{\rm L}^2    }^{(1,2)}(0,0)$}
		
		&\mathcal{Y}\left[\tiny{\young(p,r)},\tiny{\young(s,t)}\right]\left(\overline{e}_{\rm R}{}_{p} e_{\rm L}{}_{s}\right) \left(\overline{e}_{\rm R}{}_{r} e_{\rm L}{}_{t}\right)
		
		\\&\mathcal{Y}\left[\tiny{\young(pr)},\tiny{\young(st)}\right]\left(e_{\rm L}{}_{s} C e_{\rm L}{}_{t}\right) \left(\overline{e}_{\rm R}{}_{p} C \overline{e}_{\rm R}{}_{r}\right)
		
	\end{array}\\
	
	&\begin{array}{c|l}
		
		\multirow{2}*{$\mathcal{O}_{  \bar{e}_{\rm R} e_{\rm L} \nu _{\rm L}^2    }^{(1,2)}(0,2)$}
		
		&\mathcal{Y}\left[\tiny{\young(s,t)}\right]\left(\overline{e}_{\rm R}{}_{p} \nu _{\rm L}{}_{s}\right) \left(e_{\rm L}{}_{r} C \nu _{\rm L}{}_{t}\right)
		
		\\&\mathcal{Y}\left[\tiny{\young(st)}\right]\left(\nu _{\rm L}{}_{s} C \nu _{\rm L}{}_{t}\right) \left(\overline{e}_{\rm R}{}_{p} e_{\rm L}{}_{r}\right)
		
	\end{array}\\
	
	&\begin{array}{c|l}\label{cl:psiL4l}
		
		\mathcal{O}_{  \nu _{\rm L}^4    }(0,4)
		
		&\mathcal{Y}\left[\tiny{\young(pr,st)}\right]\left(\nu _{\rm L}{}_{p} C \nu _{\rm L}{}_{r}\right) \left(\nu _{\rm L}{}_{s} C \nu _{\rm L}{}_{t}\right)
		
	\end{array}
	
\end{align}

\underline{Class $  \psi {}^2 \psi^\dagger{}^2    $}: 50 types

\begin{align}
	
	&\begin{array}{c|l}\label{cl:psiL2psiR2f}
		
		\multirow{2}*{$\mathcal{O}_{  u_{\rm L}^2 \bar{u}_{\rm L} ^2    }^{(1,2)}(0,0)$}
		
		&\mathcal{Y}\left[\tiny{\young(pr)},\tiny{\young(st)}\right]\left(u_{\rm L}{}_{pa} C u_{\rm L}{}_{rb}\right) \left(\overline{u}_{\rm L}{}_{s}^{a} C \overline{u}_{\rm L}{}_{t}^{b}\right)
		
		\\&\mathcal{Y}\left[\tiny{\young(p,r)},\tiny{\young(s,t)}\right]\left(u_{\rm L}{}_{pa} C u_{\rm L}{}_{rb}\right) \left(\overline{u}_{\rm L}{}_{s}^{a} C \overline{u}_{\rm L}{}_{t}^{b}\right)
		
	\end{array}\\
	
	&\begin{array}{c|l}
		
		\mathcal{O}_{  d_{\rm R}  e_{\rm R}  u_{\rm L}^2    }(1,1)
		
		&\mathcal{Y}\left[\tiny{\young(p,r)}\right]\epsilon ^{abc} \left(u_{\rm L}{}_{pa} C u_{\rm L}{}_{rb}\right) \left(d_{\rm R}{}_{sc} C e_{\rm R}{}_{t}\right)
		
	\end{array}\\
	
	&\begin{array}{c|l}
		
		\multirow{2}*{$\mathcal{O}_{  \bar{u}_{\rm R} u_{\rm R}  u_{\rm L} \bar{u}_{\rm L}     }^{(1,2)}(0,0)$}
		
		& \left(\overline{u}_{\rm R}{}_{p}^{a} u_{\rm L}{}_{rb}\right) \left(\overline{u}_{\rm L}{}_{t}^{b} u_{\rm R}{}_{sa}\right)
		
		\\& \left(\overline{u}_{\rm R}{}_{p}^{a} u_{\rm L}{}_{ra}\right) \left(\overline{u}_{\rm L}{}_{t}^{c} u_{\rm R}{}_{sc}\right)
		
	\end{array}\\
	
	&\begin{array}{c|l}
		
		\multirow{2}*{$\mathcal{O}_{  d_{\rm R}  \bar{d}_{\rm L}  \bar{u}_{\rm R} u_{\rm L}    }^{(1,2)}(0,0)$}
		
		& \left(\overline{d}_{\rm L}{}_{t}^{b} d_{\rm R}{}_{sa}\right) \left(\overline{u}_{\rm R}{}_{p}^{a} u_{\rm L}{}_{rb}\right)
		
		\\& \left(\overline{u}_{\rm R}{}_{p}^{a} u_{\rm L}{}_{ra}\right) \left(\overline{d}_{\rm L}{}_{t}^{c} d_{\rm R}{}_{sc}\right)
		
	\end{array}\\
	
	&\begin{array}{c|l}
		
		\mathcal{O}_{  e_{\rm R}  \bar{e}_{\rm L}  \bar{u}_{\rm R} u_{\rm L}    }(0,0)
		
		& \left(\overline{e}_{\rm L}{}_{t} e_{\rm R}{}_{s}\right) \left(\overline{u}_{\rm R}{}_{p}^{a} u_{\rm L}{}_{ra}\right)
		
	\end{array}\\
	
	&\begin{array}{c|l}
		
		\mathcal{O}_{  \bar{\nu }_{\rm L} ^2 \bar{u}_{\rm R} u_{\rm L}    }(0,-2)
		
		&\mathcal{Y}\left[\tiny{\young(st)}\right]\left(\overline{\nu}_{\rm L}{}_{s} C \overline{\nu}_{\rm L}{}_{t}\right) \left(\overline{u}_{\rm R}{}_{p}^{a} u_{\rm L}{}_{ra}\right)
		
	\end{array}\\
	
	&\begin{array}{c|l}
		
		\multirow{2}*{$\mathcal{O}_{  d_{\rm L} \bar{d}_{\rm L}  u_{\rm L} \bar{u}_{\rm L}     }^{(1,2)}(0,0)$}
		
		& \left(d_{\rm L}{}_{pa} C u_{\rm L}{}_{rb}\right) \left(\overline{d}_{\rm L}{}_{s}^{a} C \overline{u}_{\rm L}{}_{t}^{b}\right)
		
		\\& \left(d_{\rm L}{}_{pa} C u_{\rm L}{}_{rb}\right) \left(\overline{d}_{\rm L}{}_{s}^{b} C \overline{u}_{\rm L}{}_{t}^{a}\right)
		
	\end{array}\\
	
	&\begin{array}{c|l}
		
		\mathcal{O}_{  d_{\rm L} e_{\rm R}  u_{\rm R}  u_{\rm L}    }(1,1)
		
		& \epsilon ^{abc} \left(e_{\rm R}{}_{s} C u_{\rm R}{}_{tc}\right) \left(d_{\rm L}{}_{pa} C u_{\rm L}{}_{rb}\right)
		
	\end{array}\\
	
	&\begin{array}{c|l}
		
		\mathcal{O}_{  d_{\rm R}  d_{\rm L} \bar{\nu }_{\rm L}  u_{\rm L}    }(1,-1)
		
		& \epsilon ^{abc} \left(\overline{\nu}_{\rm L}{}_{t} d_{\rm R}{}_{sc}\right) \left(d_{\rm L}{}_{pa} C u_{\rm L}{}_{rb}\right)
		
	\end{array}\\
	
	&\begin{array}{c|l}
		
		\multirow{2}*{$\mathcal{O}_{  \bar{d}_{\rm R} d_{\rm R}  u_{\rm L} \bar{u}_{\rm L}     }^{(1,2)}(0,0)$}
		
		& \left(\overline{d}_{\rm R}{}_{p}^{a} u_{\rm L}{}_{rb}\right) \left(\overline{u}_{\rm L}{}_{t}^{b} d_{\rm R}{}_{sa}\right)
		
		\\& \left(\overline{d}_{\rm R}{}_{p}^{a} u_{\rm L}{}_{ra}\right) \left(\overline{u}_{\rm L}{}_{t}^{c} d_{\rm R}{}_{sc}\right)
		
	\end{array}\\
	
	&\begin{array}{c|l}
		
		\mathcal{O}_{  \bar{d}_{\rm R} e_{\rm R}  \bar{\nu }_{\rm L}  u_{\rm L}    }(0,0)
		
		& \left(\overline{\nu}_{\rm L}{}_{t} e_{\rm R}{}_{s}\right) \left(\overline{d}_{\rm R}{}_{p}^{a} u_{\rm L}{}_{ra}\right)
		
	\end{array}\\
	
	&\begin{array}{c|l}
		
		\mathcal{O}_{  e_{\rm L} \bar{e}_{\rm L}  u_{\rm L} \bar{u}_{\rm L}     }(0,0)
		
		& \left(e_{\rm L}{}_{p} C u_{\rm L}{}_{ra}\right) \left(\overline{e}_{\rm L}{}_{s} C \overline{u}_{\rm L}{}_{t}^{a}\right)
		
	\end{array}\\
	
	&\begin{array}{c|l}
		
		\mathcal{O}_{  d_{\rm R}  e_{\rm L} u_{\rm R}  u_{\rm L}    }(1,1)
		
		& \epsilon ^{abc} \left(e_{\rm L}{}_{p} C u_{\rm L}{}_{ra}\right) \left(d_{\rm R}{}_{sb} C u_{\rm R}{}_{tc}\right)
		
	\end{array}\\
	
	&\begin{array}{c|l}
		
		\mathcal{O}_{  \bar{d}_{\rm L}  e_{\rm L} \bar{\nu }_{\rm L}  u_{\rm L}    }(0,0)
		
		& \left(e_{\rm L}{}_{p} C u_{\rm L}{}_{ra}\right) \left(\overline{d}_{\rm L}{}_{s}^{a} C \overline{\nu}_{\rm L}{}_{t}\right)
		
	\end{array}\\
	
	&\begin{array}{c|l}
		
		\mathcal{O}_{  \bar{e}_{\rm R} e_{\rm R}  u_{\rm L} \bar{u}_{\rm L}     }(0,0)
		
		& \left(\overline{e}_{\rm R}{}_{p} u_{\rm L}{}_{ra}\right) \left(\overline{u}_{\rm L}{}_{t}^{a} e_{\rm R}{}_{s}\right)
		
	\end{array}\\
	
	&\begin{array}{c|l}
		
		\mathcal{O}_{  \nu _{\rm L} \bar{\nu }_{\rm L}  u_{\rm L} \bar{u}_{\rm L}     }(0,0)
		
		& \left(u_{\rm L}{}_{pa} C \nu _{\rm L}{}_{r}\right) \left(\overline{u}_{\rm L}{}_{s}^{a} C \overline{\nu}_{\rm L}{}_{t}\right)
		
	\end{array}\\
	
	&\begin{array}{c|l}
		
		\mathcal{O}_{  \bar{d}_{\rm L}  e_{\rm R}  \nu _{\rm L} u_{\rm L}    }(0,2)
		
		& \left(u_{\rm L}{}_{pa} C \nu _{\rm L}{}_{r}\right) \left(\overline{d}_{\rm L}{}_{s}^{a} e_{\rm R}{}_{t}\right)
		
	\end{array}\\
	
	&\begin{array}{c|l}
		
		\mathcal{O}_{  d_{\rm R} ^2 \nu _{\rm L} u_{\rm L}    }(1,1)
		
		&\mathcal{Y}\left[\tiny{\young(s,t)}\right]\epsilon ^{abc} \left(u_{\rm L}{}_{pa} C \nu _{\rm L}{}_{r}\right) \left(d_{\rm R}{}_{sb} C d_{\rm R}{}_{tc}\right)
		
	\end{array}\\
	
	&\begin{array}{c|l}
		
		\multirow{2}*{$\mathcal{O}_{  \bar{u}_{\rm R}^2 u_{\rm R} ^2    }^{(1,2)}(0,0)$}
		
		&\mathcal{Y}\left[\tiny{\young(pr)},\tiny{\young(st)}\right]\left(u_{\rm R}{}_{sa} C u_{\rm R}{}_{tb}\right) \left(\overline{u}_{\rm R}{}_{p}^{a} C \overline{u}_{\rm R}{}_{r}^{b}\right)
		
		\\&\mathcal{Y}\left[\tiny{\young(p,r)},\tiny{\young(s,t)}\right]\left(u_{\rm R}{}_{sa} C u_{\rm R}{}_{tb}\right) \left(\overline{u}_{\rm R}{}_{p}^{a} C \overline{u}_{\rm R}{}_{r}^{b}\right)
		
	\end{array}\\
	
	&\begin{array}{c|l}
		
		\mathcal{O}_{  \bar{d}_{\rm L}  \bar{e}_{\rm L}  \bar{u}_{\rm R}^2    }(-1,-1)
		
		&\mathcal{Y}\left[\tiny{\young(p,r)}\right]\epsilon _{abc} \left(\overline{u}_{\rm R}{}_{p}^{a} C \overline{u}_{\rm R}{}_{r}^{b}\right) \left(\overline{d}_{\rm L}{}_{s}^{c} C \overline{e}_{\rm L}{}_{t}\right)
		
	\end{array}\\
	
	&\begin{array}{c|l}
		
		\multirow{2}*{$\mathcal{O}_{  d_{\rm L} \bar{d}_{\rm L}  \bar{u}_{\rm R} u_{\rm R}     }^{(1,2)}(0,0)$}
		
		& \left(\overline{u}_{\rm R}{}_{r}^{b} d_{\rm L}{}_{pa}\right) \left(\overline{d}_{\rm L}{}_{s}^{a} u_{\rm R}{}_{tb}\right)
		
		\\& \left(\overline{u}_{\rm R}{}_{r}^{a} d_{\rm L}{}_{pa}\right) \left(\overline{d}_{\rm L}{}_{s}^{c} u_{\rm R}{}_{tc}\right)
		
	\end{array}\\
	
	&\begin{array}{c|l}
		
		\mathcal{O}_{  d_{\rm L} \bar{e}_{\rm L}  \bar{\nu }_{\rm L}  \bar{u}_{\rm R}    }(0,-2)
		
		& \left(\overline{e}_{\rm L}{}_{s} C \overline{\nu}_{\rm L}{}_{t}\right) \left(\overline{u}_{\rm R}{}_{r}^{a} d_{\rm L}{}_{pa}\right)
		
	\end{array}\\
	
	&\begin{array}{c|l}
		
		\multirow{2}*{$\mathcal{O}_{  \bar{d}_{\rm R} d_{\rm R}  \bar{u}_{\rm R} u_{\rm R}     }^{(1,2)}(0,0)$}
		
		& \left(d_{\rm R}{}_{sa} C u_{\rm R}{}_{tb}\right) \left(\overline{d}_{\rm R}{}_{p}^{a} C \overline{u}_{\rm R}{}_{r}^{b}\right)
		
		\\& \left(d_{\rm R}{}_{sb} C u_{\rm R}{}_{ta}\right) \left(\overline{d}_{\rm R}{}_{p}^{a} C \overline{u}_{\rm R}{}_{r}^{b}\right)
		
	\end{array}\\
	
	&\begin{array}{c|l}
		
		\mathcal{O}_{  \bar{d}_{\rm R} \bar{d}_{\rm L}  \bar{\nu }_{\rm L}  \bar{u}_{\rm R}    }(-1,-1)
		
		& \epsilon _{abc} \left(\overline{d}_{\rm L}{}_{s}^{c} C \overline{\nu}_{\rm L}{}_{t}\right) \left(\overline{d}_{\rm R}{}_{p}^{a} C \overline{u}_{\rm R}{}_{r}^{b}\right)
		
	\end{array}\\
	
	&\begin{array}{c|l}
		
		\mathcal{O}_{  e_{\rm L} \bar{e}_{\rm L}  \bar{u}_{\rm R} u_{\rm R}     }(0,0)
		
		& \left(\overline{e}_{\rm L}{}_{s} u_{\rm R}{}_{ta}\right) \left(\overline{u}_{\rm R}{}_{r}^{a} e_{\rm L}{}_{p}\right)
		
	\end{array}\\
	
	&\begin{array}{c|l}
		
		\mathcal{O}_{  \bar{e}_{\rm R} e_{\rm R}  \bar{u}_{\rm R} u_{\rm R}     }(0,0)
		
		& \left(e_{\rm R}{}_{s} C u_{\rm R}{}_{ta}\right) \left(\overline{e}_{\rm R}{}_{p} C \overline{u}_{\rm R}{}_{r}^{a}\right)
		
	\end{array}\\
	
	&\begin{array}{c|l}
		
		\mathcal{O}_{  d_{\rm R}  \bar{e}_{\rm R} \bar{\nu }_{\rm L}  \bar{u}_{\rm R}    }(0,-2)
		
		& \left(\overline{\nu}_{\rm L}{}_{t} d_{\rm R}{}_{sa}\right) \left(\overline{e}_{\rm R}{}_{p} C \overline{u}_{\rm R}{}_{r}^{a}\right)
		
	\end{array}\\
	
	&\begin{array}{c|l}
		
		\mathcal{O}_{  \nu _{\rm L} \bar{\nu }_{\rm L}  \bar{u}_{\rm R} u_{\rm R}     }(0,0)
		
		& \left(\overline{\nu}_{\rm L}{}_{t} u_{\rm R}{}_{sa}\right) \left(\overline{u}_{\rm R}{}_{p}^{a} \nu _{\rm L}{}_{r}\right)
		
	\end{array}\\
	
	&\begin{array}{c|l}
		
		\mathcal{O}_{  \bar{d}_{\rm L} ^2 \nu _{\rm L} \bar{u}_{\rm R}    }(-1,1)
		
		&\mathcal{Y}\left[\tiny{\young(s,t)}\right]\epsilon _{abc} \left(\overline{u}_{\rm R}{}_{p}^{a} \nu _{\rm L}{}_{r}\right) \left(\overline{d}_{\rm L}{}_{s}^{b} C \overline{d}_{\rm L}{}_{t}^{c}\right)
		
	\end{array}\\
	
	&\begin{array}{c|l}
		
		\mathcal{O}_{  d_{\rm R}  \bar{e}_{\rm L}  \nu _{\rm L} \bar{u}_{\rm R}    }(0,0)
		
		& \left(\overline{e}_{\rm L}{}_{t} d_{\rm R}{}_{sa}\right) \left(\overline{u}_{\rm R}{}_{p}^{a} \nu _{\rm L}{}_{r}\right)
		
	\end{array}\\
	
	&\begin{array}{c|l}
		
		\multirow{2}*{$\mathcal{O}_{  d_{\rm L}^2 \bar{d}_{\rm L} ^2    }^{(1,2)}(0,0)$}
		
		&\mathcal{Y}\left[\tiny{\young(pr)},\tiny{\young(st)}\right]\left(d_{\rm L}{}_{pa} C d_{\rm L}{}_{rb}\right) \left(\overline{d}_{\rm L}{}_{s}^{a} C \overline{d}_{\rm L}{}_{t}^{b}\right)
		
		\\&\mathcal{Y}\left[\tiny{\young(p,r)},\tiny{\young(s,t)}\right]\left(d_{\rm L}{}_{pa} C d_{\rm L}{}_{rb}\right) \left(\overline{d}_{\rm L}{}_{s}^{a} C \overline{d}_{\rm L}{}_{t}^{b}\right)
		
	\end{array}\\
	
	&\begin{array}{c|l}
		
		\mathcal{O}_{  d_{\rm R}  d_{\rm L}^2 \bar{e}_{\rm L}     }(1,-1)
		
		&\mathcal{Y}\left[\tiny{\young(p,r)}\right]\epsilon ^{abc} \left(\overline{e}_{\rm L}{}_{t} d_{\rm R}{}_{sc}\right) \left(d_{\rm L}{}_{pa} C d_{\rm L}{}_{rb}\right)
		
	\end{array}\\
	
	&\begin{array}{c|l}
		
		\multirow{2}*{$\mathcal{O}_{  \bar{d}_{\rm R} d_{\rm R}  d_{\rm L} \bar{d}_{\rm L}     }^{(1,2)}(0,0)$}
		
		& \left(\overline{d}_{\rm R}{}_{p}^{a} d_{\rm L}{}_{rb}\right) \left(\overline{d}_{\rm L}{}_{t}^{b} d_{\rm R}{}_{sa}\right)
		
		\\& \left(\overline{d}_{\rm R}{}_{p}^{a} d_{\rm L}{}_{ra}\right) \left(\overline{d}_{\rm L}{}_{t}^{c} d_{\rm R}{}_{sc}\right)
		
	\end{array}\\
	
	&\begin{array}{c|l}
		
		\mathcal{O}_{  \bar{d}_{\rm R} d_{\rm L} e_{\rm R}  \bar{e}_{\rm L}     }(0,0)
		
		& \left(\overline{e}_{\rm L}{}_{t} e_{\rm R}{}_{s}\right) \left(\overline{d}_{\rm R}{}_{p}^{a} d_{\rm L}{}_{ra}\right)
		
	\end{array}\\
	
	&\begin{array}{c|l}
		
		\mathcal{O}_{  \bar{d}_{\rm R} d_{\rm L} \bar{\nu }_{\rm L} ^2    }(0,-2)
		
		&\mathcal{Y}\left[\tiny{\young(st)}\right]\left(\overline{\nu}_{\rm L}{}_{s} C \overline{\nu}_{\rm L}{}_{t}\right) \left(\overline{d}_{\rm R}{}_{p}^{a} d_{\rm L}{}_{ra}\right)
		
	\end{array}\\
	
	&\begin{array}{c|l}
		
		\mathcal{O}_{  d_{\rm L} \bar{d}_{\rm L}  e_{\rm L} \bar{e}_{\rm L}     }(0,0)
		
		& \left(d_{\rm L}{}_{pa} C e_{\rm L}{}_{r}\right) \left(\overline{d}_{\rm L}{}_{s}^{a} C \overline{e}_{\rm L}{}_{t}\right)
		
	\end{array}\\
	
	&\begin{array}{c|l}
		
		\mathcal{O}_{  d_{\rm L} \bar{d}_{\rm L}  \bar{e}_{\rm R} e_{\rm R}     }(0,0)
		
		& \left(\overline{e}_{\rm R}{}_{r} d_{\rm L}{}_{pa}\right) \left(\overline{d}_{\rm L}{}_{s}^{a} e_{\rm R}{}_{t}\right)
		
	\end{array}\\
	
	&\begin{array}{c|l}
		
		\mathcal{O}_{  d_{\rm R} ^2 d_{\rm L} \bar{e}_{\rm R}    }(1,-1)
		
		&\mathcal{Y}\left[\tiny{\young(s,t)}\right]\epsilon ^{abc} \left(\overline{e}_{\rm R}{}_{r} d_{\rm L}{}_{pa}\right) \left(d_{\rm R}{}_{sb} C d_{\rm R}{}_{tc}\right)
		
	\end{array}\\
	
	&\begin{array}{c|l}
		
		\mathcal{O}_{  d_{\rm L} \bar{d}_{\rm L}  \nu _{\rm L} \bar{\nu }_{\rm L}     }(0,0)
		
		& \left(d_{\rm L}{}_{pa} C \nu _{\rm L}{}_{r}\right) \left(\overline{d}_{\rm L}{}_{s}^{a} C \overline{\nu}_{\rm L}{}_{t}\right)
		
	\end{array}\\
	
	&\begin{array}{c|l}
		
		\multirow{2}*{$\mathcal{O}_{  \bar{d}_{\rm R}^2 d_{\rm R} ^2    }^{(1,2)}(0,0)$}
		
		&\mathcal{Y}\left[\tiny{\young(pr)},\tiny{\young(st)}\right]\left(d_{\rm R}{}_{sa} C d_{\rm R}{}_{tb}\right) \left(\overline{d}_{\rm R}{}_{p}^{a} C \overline{d}_{\rm R}{}_{r}^{b}\right)
		
		\\&\mathcal{Y}\left[\tiny{\young(p,r)},\tiny{\young(s,t)}\right]\left(d_{\rm R}{}_{sa} C d_{\rm R}{}_{tb}\right) \left(\overline{d}_{\rm R}{}_{p}^{a} C \overline{d}_{\rm R}{}_{r}^{b}\right)
		
	\end{array}\\
	
	&\begin{array}{c|l}
		
		\mathcal{O}_{  \bar{d}_{\rm R} d_{\rm R}  e_{\rm L} \bar{e}_{\rm L}     }(0,0)
		
		& \left(\overline{e}_{\rm L}{}_{t} d_{\rm R}{}_{sa}\right) \left(\overline{d}_{\rm R}{}_{p}^{a} e_{\rm L}{}_{r}\right)
		
	\end{array}\\
	
	&\begin{array}{c|l}
		
		\mathcal{O}_{  \bar{d}_{\rm R} d_{\rm R}  \bar{e}_{\rm R} e_{\rm R}     }(0,0)
		
		& \left(d_{\rm R}{}_{sa} C e_{\rm R}{}_{t}\right) \left(\overline{d}_{\rm R}{}_{p}^{a} C \overline{e}_{\rm R}{}_{r}\right)
		
	\end{array}\\
	
	&\begin{array}{c|l}
		
		\mathcal{O}_{  \bar{d}_{\rm R} d_{\rm R}  \nu _{\rm L} \bar{\nu }_{\rm L}     }(0,0)
		
		& \left(\overline{\nu}_{\rm L}{}_{t} d_{\rm R}{}_{sa}\right) \left(\overline{d}_{\rm R}{}_{p}^{a} \nu _{\rm L}{}_{r}\right)
		
	\end{array}\\
	
	&\begin{array}{c|l}
		
		\mathcal{O}_{  e_{\rm L}^2 \bar{e}_{\rm L} ^2    }(0,0)
		
		&\mathcal{Y}\left[\tiny{\young(pr)},\tiny{\young(st)}\right]\left(e_{\rm L}{}_{p} C e_{\rm L}{}_{r}\right) \left(\overline{e}_{\rm L}{}_{s} C \overline{e}_{\rm L}{}_{t}\right)
		
	\end{array}\\
	
	&\begin{array}{c|l}
		
		\mathcal{O}_{  \bar{e}_{\rm R} e_{\rm R}  e_{\rm L} \bar{e}_{\rm L}     }(0,0)
		
		& \left(\overline{e}_{\rm R}{}_{p} e_{\rm L}{}_{r}\right) \left(\overline{e}_{\rm L}{}_{t} e_{\rm R}{}_{s}\right)
		
	\end{array}\\
	
	&\begin{array}{c|l}
		
		\mathcal{O}_{  \bar{e}_{\rm R} e_{\rm L} \bar{\nu }_{\rm L} ^2    }(0,-2)
		
		&\mathcal{Y}\left[\tiny{\young(st)}\right]\left(\overline{\nu}_{\rm L}{}_{s} C \overline{\nu}_{\rm L}{}_{t}\right) \left(\overline{e}_{\rm R}{}_{p} e_{\rm L}{}_{r}\right)
		
	\end{array}\\
	
	&\begin{array}{c|l}
		
		\mathcal{O}_{  e_{\rm L} \bar{e}_{\rm L}  \nu _{\rm L} \bar{\nu }_{\rm L}     }(0,0)
		
		& \left(e_{\rm L}{}_{p} C \nu _{\rm L}{}_{r}\right) \left(\overline{e}_{\rm L}{}_{s} C \overline{\nu}_{\rm L}{}_{t}\right)
		
	\end{array}\\
	
	&\begin{array}{c|l}
		
		\mathcal{O}_{  \bar{e}_{\rm R}^2 e_{\rm R} ^2    }(0,0)
		
		&\mathcal{Y}\left[\tiny{\young(pr)},\tiny{\young(st)}\right]\left(e_{\rm R}{}_{s} C e_{\rm R}{}_{t}\right) \left(\overline{e}_{\rm R}{}_{p} C \overline{e}_{\rm R}{}_{r}\right)
		
	\end{array}\\
	
	&\begin{array}{c|l}
		
		\mathcal{O}_{  \bar{e}_{\rm R} e_{\rm R}  \nu _{\rm L} \bar{\nu }_{\rm L}     }(0,0)
		
		& \left(\overline{e}_{\rm R}{}_{p} \nu _{\rm L}{}_{r}\right) \left(\overline{\nu}_{\rm L}{}_{t} e_{\rm R}{}_{s}\right)
		
	\end{array}\\
	
	&\begin{array}{c|l}\label{cl:psiL2psiR2l}
		
		\mathcal{O}_{  \nu _{\rm L}^2 \bar{\nu }_{\rm L} ^2    }(0,0)
		
		&\mathcal{Y}\left[\tiny{\young(pr)},\tiny{\young(st)}\right]\left(\nu _{\rm L}{}_{p} C \nu _{\rm L}{}_{r}\right) \left(\overline{\nu}_{\rm L}{}_{s} C \overline{\nu}_{\rm L}{}_{t}\right)
		
	\end{array}
	
\end{align}

\subsection{Lists of the Dim-7 Operators}

\indent \indent \underline{Class $F_{\rm{L}}{}^2 \psi {}^2    $}: 10 types

\begin{align}
	
	&\begin{array}{c|l}\label{cl:FL2psiL2f}
		
		\multirow{3}*{$\mathcal{O}_{G_{\rm{L}}{}^2 \bar{u}_{\rm R} u_{\rm L}    }^{(1\sim 3)}(0,0)$}
		
		& G_{\rm{L}}^{A}{}_{\mu }{}_{\nu } G_{\rm{L}}^{A}{}^{\mu }{}^{\nu } \left(\overline{u}_{\rm R}{}_{p}^{a} u_{\rm L}{}_{ra}\right)
		
		\\& d^{ABC} \left(\lambda ^C\right)_a^b G_{\rm{L}}^{A}{}_{\mu }{}_{\nu } G_{\rm{L}}^{B}{}^{\mu }{}^{\nu } \left(\overline{u}_{\rm R}{}_{p}^{a} u_{\rm L}{}_{rb}\right)
		
		\\& i f^{ABC} \left(\lambda ^C\right)_a^b G_{\rm{L}}^{A}{}_{\mu }{}_{\nu } G_{\rm{L}}^{B}{}^{\mu }{}{}_{\lambda } \left(\overline{u}_{\rm R}{}_{p}^{a} \sigma ^{\nu }{}^{\lambda } u_{\rm L}{}_{rb}\right)
		
	\end{array}\\
	
	&\begin{array}{c|l}
		
		\multirow{3}*{$\mathcal{O}_{G_{\rm{L}}{}^2 \bar{d}_{\rm R} d_{\rm L}    }^{(1\sim 3)}(0,0)$}
		
		& G_{\rm{L}}^{A}{}_{\mu }{}_{\nu } G_{\rm{L}}^{A}{}^{\mu }{}^{\nu } \left(\overline{d}_{\rm R}{}_{p}^{a} d_{\rm L}{}_{ra}\right)
		
		\\& d^{ABC} \left(\lambda ^C\right)_a^b G_{\rm{L}}^{A}{}_{\mu }{}_{\nu } G_{\rm{L}}^{B}{}^{\mu }{}^{\nu } \left(\overline{d}_{\rm R}{}_{p}^{a} d_{\rm L}{}_{rb}\right)
		
		\\& i f^{ABC} \left(\lambda ^C\right)_a^b G_{\rm{L}}^{A}{}_{\mu }{}_{\nu } G_{\rm{L}}^{B}{}^{\mu }{}{}_{\lambda } \left(\overline{d}_{\rm R}{}_{p}^{a} \sigma ^{\nu }{}^{\lambda } d_{\rm L}{}_{rb}\right)
		
	\end{array}\\
	
	&\begin{array}{c|l}
		
		\mathcal{O}_{G_{\rm{L}}{}^2 \bar{e}_{\rm R} e_{\rm L}    }(0,0)
		
		& \left(\overline{e}_{\rm R}{}_{p} e_{\rm L}{}_{r}\right) G_{\rm{L}}^{A}{}_{\mu }{}_{\nu } G_{\rm{L}}^{A}{}^{\mu }{}^{\nu }
		
	\end{array}\\
	
	&\begin{array}{c|l}
		
		\mathcal{O}_{G_{\rm{L}}{}^2 \nu _{\rm L}^2    }(0,2)
		
		&\mathcal{Y}\left[\tiny{\young(pr)}\right]\left(\nu _{\rm L}{}_{p} C \nu _{\rm L}{}_{r}\right) G_{\rm{L}}^{A}{}_{\mu }{}_{\nu } G_{\rm{L}}^{A}{}^{\mu }{}^{\nu }
		
	\end{array}\\
	
	&\begin{array}{c|l}
		
		\multirow{2}*{$\mathcal{O}_{F_{\rm{L}} G_{\rm{L}} \bar{u}_{\rm R} u_{\rm L}    }^{(1,2)}(0,0)$}
		
		& \left(\lambda ^A\right)_a^b F_{\rm{L}}{}_{\mu }{}_{\nu } G_{\rm{L}}^{A}{}^{\mu }{}^{\nu } \left(\overline{u}_{\rm R}{}_{p}^{a} u_{\rm L}{}_{rb}\right)
		
		\\& i \left(\lambda ^A\right)_a^b F_{\rm{L}}{}_{\mu }{}_{\nu } G_{\rm{L}}^{A}{}^{\mu }{}{}_{\lambda } \left(\overline{u}_{\rm R}{}_{p}^{a} \sigma ^{\nu }{}^{\lambda } u_{\rm L}{}_{rb}\right)
		
	\end{array}\\
	
	&\begin{array}{c|l}
		
		\multirow{2}*{$\mathcal{O}_{F_{\rm{L}} G_{\rm{L}} \bar{d}_{\rm R} d_{\rm L}    }^{(1,2)}(0,0)$}
		
		& \left(\lambda ^A\right)_a^b F_{\rm{L}}{}_{\mu }{}_{\nu } G_{\rm{L}}^{A}{}^{\mu }{}^{\nu } \left(\overline{d}_{\rm R}{}_{p}^{a} d_{\rm L}{}_{rb}\right)
		
		\\& i \left(\lambda ^A\right)_a^b F_{\rm{L}}{}_{\mu }{}_{\nu } G_{\rm{L}}^{A}{}^{\mu }{}{}_{\lambda } \left(\overline{d}_{\rm R}{}_{p}^{a} \sigma ^{\nu }{}^{\lambda } d_{\rm L}{}_{rb}\right)
		
	\end{array}\\
	
	&\begin{array}{c|l}
		
		\mathcal{O}_{F_{\rm{L}}{}^2 \bar{u}_{\rm R} u_{\rm L}    }(0,0)
		
		& F_{\rm{L}}{}_{\mu }{}_{\nu } F_{\rm{L}}{}^{\mu }{}^{\nu } \left(\overline{u}_{\rm R}{}_{p}^{a} u_{\rm L}{}_{ra}\right)
		
	\end{array}\\
	
	&\begin{array}{c|l}
		
		\mathcal{O}_{F_{\rm{L}}{}^2 \bar{d}_{\rm R} d_{\rm L}    }(0,0)
		
		& F_{\rm{L}}{}_{\mu }{}_{\nu } F_{\rm{L}}{}^{\mu }{}^{\nu } \left(\overline{d}_{\rm R}{}_{p}^{a} d_{\rm L}{}_{ra}\right)
		
	\end{array}\\
	
	&\begin{array}{c|l}
		
		\mathcal{O}_{F_{\rm{L}}{}^2 \bar{e}_{\rm R} e_{\rm L}    }(0,0)
		
		& F_{\rm{L}}{}_{\mu }{}_{\nu } F_{\rm{L}}{}^{\mu }{}^{\nu } \left(\overline{e}_{\rm R}{}_{p} e_{\rm L}{}_{r}\right)
		
	\end{array}\\
	
	&\begin{array}{c|l}\label{cl:FL2psiL2l}
		
		\mathcal{O}_{F_{\rm{L}}{}^2 \nu _{\rm L}^2    }(0,2)
		
		&\mathcal{Y}\left[\tiny{\young(pr)}\right]F_{\rm{L}}{}_{\mu }{}_{\nu } \left(\nu _{\rm L}{}_{p} C \nu _{\rm L}{}_{r}\right) F_{\rm{L}}{}^{\mu }{}^{\nu }
		
	\end{array}
	
\end{align}

\underline{Class $F_{\rm{L}}{}^2 \psi^\dagger{}^2    $}: 10 types

\begin{align}
	
	&\begin{array}{c|l}\label{cl:FL2psiR2f}
		
		\multirow{2}*{$\mathcal{O}_{G_{\rm{L}}{}^2 u_{\rm R}  \bar{u}_{\rm L}     }^{(1,2)}(0,0)$}
		
		& G_{\rm{L}}^{A}{}_{\mu }{}_{\nu } G_{\rm{L}}^{A}{}^{\mu }{}^{\nu } \left(\overline{u}_{\rm L}{}_{r}^{a} u_{\rm R}{}_{pa}\right)
		
		\\& d^{ABC} \left(\lambda ^C\right)_b^a G_{\rm{L}}^{A}{}_{\mu }{}_{\nu } G_{\rm{L}}^{B}{}^{\mu }{}^{\nu } \left(\overline{u}_{\rm L}{}_{r}^{b} u_{\rm R}{}_{pa}\right)
		
	\end{array}\\
	
	&\begin{array}{c|l}
		
		\multirow{2}*{$\mathcal{O}_{G_{\rm{L}}{}^2 d_{\rm R}  \bar{d}_{\rm L}     }^{(1,2)}(0,0)$}
		
		& G_{\rm{L}}^{A}{}_{\mu }{}_{\nu } G_{\rm{L}}^{A}{}^{\mu }{}^{\nu } \left(\overline{d}_{\rm L}{}_{r}^{a} d_{\rm R}{}_{pa}\right)
		
		\\& d^{ABC} \left(\lambda ^C\right)_b^a G_{\rm{L}}^{A}{}_{\mu }{}_{\nu } G_{\rm{L}}^{B}{}^{\mu }{}^{\nu } \left(\overline{d}_{\rm L}{}_{r}^{b} d_{\rm R}{}_{pa}\right)
		
	\end{array}\\
	
	&\begin{array}{c|l}
		
		\mathcal{O}_{G_{\rm{L}}{}^2 e_{\rm R}  \bar{e}_{\rm L}     }(0,0)
		
		& \left(\overline{e}_{\rm L}{}_{r} e_{\rm R}{}_{p}\right) G_{\rm{L}}^{A}{}_{\mu }{}_{\nu } G_{\rm{L}}^{A}{}^{\mu }{}^{\nu }
		
	\end{array}\\
	
	&\begin{array}{c|l}
		
		\mathcal{O}_{G_{\rm{L}}{}^2 \bar{\nu }_{\rm L} ^2    }(0,-2)
		
		&\mathcal{Y}\left[\tiny{\young(pr)}\right]\left(\overline{\nu}_{\rm L}{}_{p} C \overline{\nu}_{\rm L}{}_{r}\right) G_{\rm{L}}^{A}{}_{\mu }{}_{\nu } G_{\rm{L}}^{A}{}^{\mu }{}^{\nu }
		
	\end{array}\\
	
	&\begin{array}{c|l}
		
		\mathcal{O}_{F_{\rm{L}} G_{\rm{L}} u_{\rm R}  \bar{u}_{\rm L}     }(0,0)
		
		& \left(\lambda ^A\right)_b^a F_{\rm{L}}{}_{\mu }{}_{\nu } G_{\rm{L}}^{A}{}^{\mu }{}^{\nu } \left(\overline{u}_{\rm L}{}_{r}^{b} u_{\rm R}{}_{pa}\right)
		
	\end{array}\\
	
	&\begin{array}{c|l}
		
		\mathcal{O}_{F_{\rm{L}} G_{\rm{L}} d_{\rm R}  \bar{d}_{\rm L}     }(0,0)
		
		& \left(\lambda ^A\right)_b^a F_{\rm{L}}{}_{\mu }{}_{\nu } G_{\rm{L}}^{A}{}^{\mu }{}^{\nu } \left(\overline{d}_{\rm L}{}_{r}^{b} d_{\rm R}{}_{pa}\right)
		
	\end{array}\\
	
	&\begin{array}{c|l}
		
		\mathcal{O}_{F_{\rm{L}}{}^2 u_{\rm R}  \bar{u}_{\rm L}     }(0,0)
		
		& F_{\rm{L}}{}_{\mu }{}_{\nu } F_{\rm{L}}{}^{\mu }{}^{\nu } \left(\overline{u}_{\rm L}{}_{r}^{a} u_{\rm R}{}_{pa}\right)
		
	\end{array}\\
	
	&\begin{array}{c|l}
		
		\mathcal{O}_{F_{\rm{L}}{}^2 d_{\rm R}  \bar{d}_{\rm L}     }(0,0)
		
		& F_{\rm{L}}{}_{\mu }{}_{\nu } F_{\rm{L}}{}^{\mu }{}^{\nu } \left(\overline{d}_{\rm L}{}_{r}^{a} d_{\rm R}{}_{pa}\right)
		
	\end{array}\\
	
	&\begin{array}{c|l}
		
		\mathcal{O}_{F_{\rm{L}}{}^2 e_{\rm R}  \bar{e}_{\rm L}     }(0,0)
		
		& F_{\rm{L}}{}_{\mu }{}_{\nu } F_{\rm{L}}{}^{\mu }{}^{\nu } \left(\overline{e}_{\rm L}{}_{r} e_{\rm R}{}_{p}\right)
		
	\end{array}\\
	
	&\begin{array}{c|l}\label{cl:FL2psiR2l}
		
		\mathcal{O}_{F_{\rm{L}}{}^2 \bar{\nu }_{\rm L} ^2    }(0,-2)
		
		&\mathcal{Y}\left[\tiny{\young(pr)}\right]F_{\rm{L}}{}_{\mu }{}_{\nu } F_{\rm{L}}{}^{\mu }{}^{\nu } \left(\overline{\nu}_{\rm L}{}_{p} C \overline{\nu}_{\rm L}{}_{r}\right)
		
	\end{array}
	
\end{align}

\underline{Class $  \psi {}^3 \psi^\dagger   D$}: 52 types

\begin{align}
	
	&\begin{array}{c|l}\label{cl:psiL3psiRDf}
		
		\multirow{2}*{$\mathcal{O}_{  \bar{u}_{\rm R} u_{\rm L}^2 \bar{u}_{\rm L}    D}^{(1,2)}(0,0)$}
		
		&\mathcal{Y}\left[\tiny{\young(r,s)}\right]i \left(u_{\rm L}{}_{rb} C D^{\mu } u_{\rm L}{}_{sa}\right) \left(\overline{u}_{\rm R}{}_{p}^{a} \gamma _{\mu } C \overline{u}_{\rm L}{}_{t}^{b}\right)
		
		\\&\mathcal{Y}\left[\tiny{\young(rs)}\right]i \left(u_{\rm L}{}_{rb} C D^{\mu } u_{\rm L}{}_{sa}\right) \left(\overline{u}_{\rm R}{}_{p}^{a} \gamma _{\mu } C \overline{u}_{\rm L}{}_{t}^{b}\right)
		
	\end{array}\\
	
	&\begin{array}{c|l}
		
		\mathcal{O}_{  d_{\rm L} e_{\rm R}  u_{\rm L}^2   D}(1,1)
		
		&\mathcal{Y}\left[\tiny{\young(rs)}\right]i \epsilon ^{abc} \left(d_{\rm L}{}_{pa} C \gamma _{\mu } e_{\rm R}{}_{t}\right) \left(u_{\rm L}{}_{rb} C D^{\mu } u_{\rm L}{}_{sc}\right)
		
	\end{array}\\
	
	&\begin{array}{c|l}
		
		\mathcal{O}_{  d_{\rm R}  e_{\rm L} u_{\rm L}^2   D}(1,1)
		
		&\mathcal{Y}\left[\tiny{\young(rs)}\right]i \epsilon ^{abc} \left(e_{\rm L}{}_{p} C \gamma _{\mu } d_{\rm R}{}_{tc}\right) \left(u_{\rm L}{}_{ra} C D^{\mu } u_{\rm L}{}_{sb}\right)
		
	\end{array}\\
	
	&\begin{array}{c|l}
		
		\multirow{2}*{$\mathcal{O}_{  \bar{u}_{\rm R}^2 u_{\rm R}  u_{\rm L}   D}^{(1,2)}(0,0)$}
		
		&\mathcal{Y}\left[\tiny{\young(p,r)}\right]i \left(\overline{u}_{\rm R}{}_{p}^{a} \gamma _{\mu } u_{\rm R}{}_{tb}\right) \left(\overline{u}_{\rm R}{}_{r}^{b} D^{\mu } u_{\rm L}{}_{sa}\right)
		
		\\&\mathcal{Y}\left[\tiny{\young(pr)}\right]i \left(\overline{u}_{\rm R}{}_{p}^{a} \gamma _{\mu } u_{\rm R}{}_{tb}\right) \left(\overline{u}_{\rm R}{}_{r}^{b} D^{\mu } u_{\rm L}{}_{sa}\right)
		
	\end{array}\\
	
	&\begin{array}{c|l}
		
		\multirow{2}*{$\mathcal{O}_{  d_{\rm L} \bar{d}_{\rm L}  \bar{u}_{\rm R} u_{\rm L}   D}^{(1,2)}(0,0)$}
		
		& i \left(\overline{d}_{\rm L}{}_{t}^{a} \gamma _{\mu } d_{\rm L}{}_{pa}\right) \left(\overline{u}_{\rm R}{}_{r}^{b} D^{\mu } u_{\rm L}{}_{sb}\right)
		
		\\& i \left(\overline{d}_{\rm L}{}_{t}^{c} \gamma _{\mu } d_{\rm L}{}_{pa}\right) \left(\overline{u}_{\rm R}{}_{r}^{a} D^{\mu } u_{\rm L}{}_{sc}\right)
		
	\end{array}\\
	
	&\begin{array}{c|l}
		
		\multirow{2}*{$\mathcal{O}_{  \bar{d}_{\rm R} d_{\rm R}  \bar{u}_{\rm R} u_{\rm L}   D}^{(1,2)}(0,0)$}
		
		& i \left(\overline{d}_{\rm R}{}_{p}^{a} \gamma _{\mu } d_{\rm R}{}_{tb}\right) \left(\overline{u}_{\rm R}{}_{r}^{b} D^{\mu } u_{\rm L}{}_{sa}\right)
		
		\\& i \left(\overline{d}_{\rm R}{}_{p}^{a} \gamma _{\mu } d_{\rm R}{}_{ta}\right) \left(\overline{u}_{\rm R}{}_{r}^{b} D^{\mu } u_{\rm L}{}_{sb}\right)
		
	\end{array}\\
	
	&\begin{array}{c|l}
		
		\mathcal{O}_{  e_{\rm L} \bar{e}_{\rm L}  \bar{u}_{\rm R} u_{\rm L}   D}(0,0)
		
		& i \left(\overline{e}_{\rm L}{}_{t} \gamma _{\mu } e_{\rm L}{}_{p}\right) \left(\overline{u}_{\rm R}{}_{r}^{a} D^{\mu } u_{\rm L}{}_{sa}\right)
		
	\end{array}\\
	
	&\begin{array}{c|l}
		
		\mathcal{O}_{  \bar{e}_{\rm R} e_{\rm R}  \bar{u}_{\rm R} u_{\rm L}   D}(0,0)
		
		& i \left(\overline{e}_{\rm R}{}_{p} \gamma _{\mu } e_{\rm R}{}_{t}\right) \left(\overline{u}_{\rm R}{}_{r}^{a} D^{\mu } u_{\rm L}{}_{sa}\right)
		
	\end{array}\\
	
	&\begin{array}{c|l}
		
		\mathcal{O}_{  \nu _{\rm L} \bar{\nu }_{\rm L}  \bar{u}_{\rm R} u_{\rm L}   D}(0,0)
		
		& i \left(u_{\rm L}{}_{ra} C D^{\mu } \nu _{\rm L}{}_{s}\right) \left(\overline{u}_{\rm R}{}_{p}^{a} \gamma _{\mu } C \overline{\nu}_{\rm L}{}_{t}\right)
		
	\end{array}\\
	
	&\begin{array}{c|l}
		
		\mathcal{O}_{  d_{\rm L}^2 \bar{\nu }_{\rm L}  u_{\rm L}   D}(1,-1)
		
		&\mathcal{Y}\left[\tiny{\young(pr)}\right]i \epsilon ^{abc} \left(\overline{\nu}_{\rm L}{}_{t} \gamma _{\mu } d_{\rm L}{}_{pa}\right) \left(d_{\rm L}{}_{rb} C D^{\mu } u_{\rm L}{}_{sc}\right)
		
	\end{array}\\
	
	&\begin{array}{c|l}
		
		\multirow{2}*{$\mathcal{O}_{  \bar{d}_{\rm R} d_{\rm L} u_{\rm L} \bar{u}_{\rm L}    D}^{(1,2)}(0,0)$}
		
		& i \left(d_{\rm L}{}_{rb} C D^{\mu } u_{\rm L}{}_{sa}\right) \left(\overline{d}_{\rm R}{}_{p}^{a} \gamma _{\mu } C \overline{u}_{\rm L}{}_{t}^{b}\right)
		
		\\& i \left(d_{\rm L}{}_{ra} C D^{\mu } u_{\rm L}{}_{sc}\right) \left(\overline{d}_{\rm R}{}_{p}^{a} \gamma _{\mu } C \overline{u}_{\rm L}{}_{t}^{c}\right)
		
	\end{array}\\
	
	&\begin{array}{c|l}
		
		\mathcal{O}_{  d_{\rm L} e_{\rm L} u_{\rm R}  u_{\rm L}   D}(1,1)
		
		& i \epsilon ^{abc} \left(d_{\rm L}{}_{pa} C \gamma _{\mu } u_{\rm R}{}_{tc}\right) \left(e_{\rm L}{}_{r} C D^{\mu } u_{\rm L}{}_{sb}\right)
		
	\end{array}\\
	
	&\begin{array}{c|l}
		
		\mathcal{O}_{  d_{\rm R}  d_{\rm L} \nu _{\rm L} u_{\rm L}   D}(1,1)
		
		& i \epsilon ^{abc} \left(d_{\rm L}{}_{pa} C \gamma _{\mu } d_{\rm R}{}_{tc}\right) \left(u_{\rm L}{}_{rb} C D^{\mu } \nu _{\rm L}{}_{s}\right)
		
	\end{array}\\
	
	&\begin{array}{c|l}
		
		\mathcal{O}_{  \bar{d}_{\rm R} e_{\rm L} \bar{\nu }_{\rm L}  u_{\rm L}   D}(0,0)
		
		& i \left(e_{\rm L}{}_{r} C D^{\mu } u_{\rm L}{}_{sa}\right) \left(\overline{d}_{\rm R}{}_{p}^{a} \gamma _{\mu } C \overline{\nu}_{\rm L}{}_{t}\right)
		
	\end{array}\\
	
	&\begin{array}{c|l}
		
		\mathcal{O}_{  \bar{d}_{\rm R} e_{\rm R}  \nu _{\rm L} u_{\rm L}   D}(0,2)
		
		& i \left(\overline{d}_{\rm R}{}_{p}^{a} \gamma _{\mu } e_{\rm R}{}_{t}\right) \left(u_{\rm L}{}_{ra} C D^{\mu } \nu _{\rm L}{}_{s}\right)
		
	\end{array}\\
	
	&\begin{array}{c|l}
		
		\mathcal{O}_{  \bar{e}_{\rm R} e_{\rm L} u_{\rm L} \bar{u}_{\rm L}    D}(0,0)
		
		& i \left(e_{\rm L}{}_{r} C D^{\mu } u_{\rm L}{}_{sa}\right) \left(\overline{e}_{\rm R}{}_{p} \gamma _{\mu } C \overline{u}_{\rm L}{}_{t}^{a}\right)
		
	\end{array}\\
	
	&\begin{array}{c|l}
		
		\mathcal{O}_{  \bar{d}_{\rm L}  e_{\rm L} \nu _{\rm L} u_{\rm L}   D}(0,2)
		
		& i \left(\overline{d}_{\rm L}{}_{t}^{a} \gamma _{\mu } e_{\rm L}{}_{p}\right) \left(u_{\rm L}{}_{ra} C D^{\mu } \nu _{\rm L}{}_{s}\right)
		
	\end{array}\\
	
	&\begin{array}{c|l}
		
		\mathcal{O}_{  \nu _{\rm L}^2 u_{\rm L} \bar{u}_{\rm L}    D}(0,2)
		
		&\mathcal{Y}\left[\tiny{\young(r,s)}\right]i \left(\nu _{\rm L}{}_{r} C D^{\mu } \nu _{\rm L}{}_{s}\right) \left(\overline{u}_{\rm L}{}_{t}^{a} \gamma _{\mu } u_{\rm L}{}_{pa}\right)
		
	\end{array}\\
	
	&\begin{array}{c|l}
		
		\mathcal{O}_{  \bar{d}_{\rm R} \bar{e}_{\rm L}  \bar{u}_{\rm R}^2   D}(-1,-1)
		
		&\mathcal{Y}\left[\tiny{\young(rs)}\right]i \epsilon _{abc} \left(\overline{d}_{\rm R}{}_{p}^{a} \gamma _{\mu } C \overline{e}_{\rm L}{}_{t}\right) \left(\overline{u}_{\rm R}{}_{r}^{b} C D^{\mu } \overline{u}_{\rm R}{}_{s}^{c}\right)
		
	\end{array}\\
	
	&\begin{array}{c|l}
		
		\mathcal{O}_{  \bar{d}_{\rm L}  \bar{e}_{\rm R} \bar{u}_{\rm R}^2   D}(-1,-1)
		
		&\mathcal{Y}\left[\tiny{\young(rs)}\right]i \epsilon _{abc} \left(\overline{u}_{\rm R}{}_{r}^{a} C D^{\mu } \overline{u}_{\rm R}{}_{s}^{b}\right) \left(\overline{e}_{\rm R}{}_{p} \gamma _{\mu } C \overline{d}_{\rm L}{}_{t}^{c}\right)
		
	\end{array}\\
	
	&\begin{array}{c|l}
		
		\multirow{2}*{$\mathcal{O}_{  \bar{d}_{\rm R} d_{\rm L} \bar{u}_{\rm R} u_{\rm R}    D}^{(1,2)}(0,0)$}
		
		& i \left(\overline{d}_{\rm R}{}_{p}^{a} \gamma _{\mu } u_{\rm R}{}_{ta}\right) \left(D^{\mu } \overline{u}_{\rm R}{}_{s}^{b} d_{\rm L}{}_{rb}\right)
		
		\\& i \left(D^{\mu } \overline{u}_{\rm R}{}_{s}^{c} d_{\rm L}{}_{ra}\right) \left(\overline{d}_{\rm R}{}_{p}^{a} \gamma _{\mu } u_{\rm R}{}_{tc}\right)
		
	\end{array}\\
	
	&\begin{array}{c|l}
		
		\mathcal{O}_{  d_{\rm L} \bar{e}_{\rm R} \bar{\nu }_{\rm L}  \bar{u}_{\rm R}   D}(0,-2)
		
		& i \left(\overline{\nu}_{\rm L}{}_{t} \gamma _{\mu } d_{\rm L}{}_{pa}\right) \left(\overline{e}_{\rm R}{}_{r} C D^{\mu } \overline{u}_{\rm R}{}_{s}^{a}\right)
		
	\end{array}\\
	
	&\begin{array}{c|l}
		
		\mathcal{O}_{  d_{\rm L} \bar{e}_{\rm L}  \nu _{\rm L} \bar{u}_{\rm R}   D}(0,0)
		
		& i \left(\overline{e}_{\rm L}{}_{t} \gamma _{\mu } d_{\rm L}{}_{pa}\right) \left(\overline{u}_{\rm R}{}_{r}^{a} D^{\mu } \nu _{\rm L}{}_{s}\right)
		
	\end{array}\\
	
	&\begin{array}{c|l}
		
		\mathcal{O}_{  \bar{d}_{\rm R}^2 \bar{\nu }_{\rm L}  \bar{u}_{\rm R}   D}(-1,-1)
		
		&\mathcal{Y}\left[\tiny{\young(pr)}\right]i \epsilon _{abc} \left(\overline{d}_{\rm R}{}_{p}^{a} \gamma _{\mu } C \overline{\nu}_{\rm L}{}_{t}\right) \left(\overline{d}_{\rm R}{}_{r}^{b} C D^{\mu } \overline{u}_{\rm R}{}_{s}^{c}\right)
		
	\end{array}\\
	
	&\begin{array}{c|l}
		
		\mathcal{O}_{  \bar{d}_{\rm R} \bar{e}_{\rm R} \bar{u}_{\rm R} \bar{u}_{\rm L}    D}(-1,-1)
		
		& i \epsilon _{abc} \left(\overline{e}_{\rm R}{}_{r} C D^{\mu } \overline{u}_{\rm R}{}_{s}^{b}\right) \left(\overline{d}_{\rm R}{}_{p}^{a} \gamma _{\mu } C \overline{u}_{\rm L}{}_{t}^{c}\right)
		
	\end{array}\\
	
	&\begin{array}{c|l}
		
		\mathcal{O}_{  \bar{d}_{\rm R} \bar{d}_{\rm L}  \nu _{\rm L} \bar{u}_{\rm R}   D}(-1,1)
		
		& i \epsilon _{abc} \left(\overline{u}_{\rm R}{}_{r}^{b} D^{\mu } \nu _{\rm L}{}_{s}\right) \left(\overline{d}_{\rm R}{}_{p}^{a} \gamma _{\mu } C \overline{d}_{\rm L}{}_{t}^{c}\right)
		
	\end{array}\\
	
	&\begin{array}{c|l}
		
		\mathcal{O}_{  \bar{e}_{\rm R} e_{\rm L} \bar{u}_{\rm R} u_{\rm R}    D}(0,0)
		
		& i \left(\overline{e}_{\rm R}{}_{p} \gamma _{\mu } u_{\rm R}{}_{ta}\right) \left(D^{\mu } \overline{u}_{\rm R}{}_{s}^{a} e_{\rm L}{}_{r}\right)
		
	\end{array}\\
	
	&\begin{array}{c|l}
		
		\mathcal{O}_{  d_{\rm R}  \bar{e}_{\rm R} \nu _{\rm L} \bar{u}_{\rm R}   D}(0,0)
		
		& i \left(\overline{e}_{\rm R}{}_{p} \gamma _{\mu } d_{\rm R}{}_{ta}\right) \left(\overline{u}_{\rm R}{}_{r}^{a} D^{\mu } \nu _{\rm L}{}_{s}\right)
		
	\end{array}\\
	
	&\begin{array}{c|l}
		
		\mathcal{O}_{  \nu _{\rm L}^2 \bar{u}_{\rm R} u_{\rm R}    D}(0,2)
		
		&\mathcal{Y}\left[\tiny{\young(r,s)}\right]i \left(\nu _{\rm L}{}_{r} C D^{\mu } \nu _{\rm L}{}_{s}\right) \left(\overline{u}_{\rm R}{}_{p}^{a} \gamma _{\mu } u_{\rm R}{}_{ta}\right)
		
	\end{array}\\
	
	&\begin{array}{c|l}
		
		\mathcal{O}_{  d_{\rm L}^3 \bar{e}_{\rm L}    D}(1,-1)
		
		&\mathcal{Y}\left[\tiny{\young(prs)}\right]i \epsilon ^{abc} \left(\overline{e}_{\rm L}{}_{t} \gamma _{\mu } d_{\rm L}{}_{pa}\right) \left(d_{\rm L}{}_{rb} C D^{\mu } d_{\rm L}{}_{sc}\right)
		
	\end{array}\\
	
	&\begin{array}{c|l}
		
		\multirow{2}*{$\mathcal{O}_{  \bar{d}_{\rm R} d_{\rm L}^2 \bar{d}_{\rm L}    D}^{(1,2)}(0,0)$}
		
		&\mathcal{Y}\left[\tiny{\young(r,s)}\right]i \left(d_{\rm L}{}_{rb} C D^{\mu } d_{\rm L}{}_{sa}\right) \left(\overline{d}_{\rm R}{}_{p}^{a} \gamma _{\mu } C \overline{d}_{\rm L}{}_{t}^{b}\right)
		
		\\&\mathcal{Y}\left[\tiny{\young(rs)}\right]i \left(d_{\rm L}{}_{rb} C D^{\mu } d_{\rm L}{}_{sa}\right) \left(\overline{d}_{\rm R}{}_{p}^{a} \gamma _{\mu } C \overline{d}_{\rm L}{}_{t}^{b}\right)
		
	\end{array}\\
	
	&\begin{array}{c|l}
		
		\mathcal{O}_{  d_{\rm R}  d_{\rm L}^2 \bar{e}_{\rm R}   D}(1,-1)
		
		&\mathcal{Y}\left[\tiny{\young(pr)}\right]i \epsilon ^{abc} \left(D^{\mu } \overline{e}_{\rm R}{}_{s} d_{\rm L}{}_{rb}\right) \left(d_{\rm L}{}_{pa} C \gamma _{\mu } d_{\rm R}{}_{tc}\right)
		
	\end{array}\\
	
	&\begin{array}{c|l}
		
		\mathcal{O}_{  d_{\rm L}^2 \nu _{\rm L} u_{\rm R}    D}(1,1)
		
		&\mathcal{Y}\left[\tiny{\young(pr)}\right]i \epsilon ^{abc} \left(d_{\rm L}{}_{pa} C \gamma _{\mu } u_{\rm R}{}_{tc}\right) \left(d_{\rm L}{}_{rb} C D^{\mu } \nu _{\rm L}{}_{s}\right)
		
	\end{array}\\
	
	&\begin{array}{c|l}
		
		\multirow{2}*{$\mathcal{O}_{  \bar{d}_{\rm R}^2 d_{\rm R}  d_{\rm L}   D}^{(1,2)}(0,0)$}
		
		&\mathcal{Y}\left[\tiny{\young(p,r)}\right]i \left(\overline{d}_{\rm R}{}_{p}^{a} \gamma _{\mu } d_{\rm R}{}_{tb}\right) \left(\overline{d}_{\rm R}{}_{r}^{b} D^{\mu } d_{\rm L}{}_{sa}\right)
		
		\\&\mathcal{Y}\left[\tiny{\young(pr)}\right]i \left(\overline{d}_{\rm R}{}_{p}^{a} \gamma _{\mu } d_{\rm R}{}_{tb}\right) \left(\overline{d}_{\rm R}{}_{r}^{b} D^{\mu } d_{\rm L}{}_{sa}\right)
		
	\end{array}\\
	
	&\begin{array}{c|l}
		
		\mathcal{O}_{  \bar{d}_{\rm R} d_{\rm L} e_{\rm L} \bar{e}_{\rm L}    D}(0,0)
		
		& i \left(d_{\rm L}{}_{ra} C D^{\mu } e_{\rm L}{}_{s}\right) \left(\overline{d}_{\rm R}{}_{p}^{a} \gamma _{\mu } C \overline{e}_{\rm L}{}_{t}\right)
		
	\end{array}\\
	
	&\begin{array}{c|l}
		
		\mathcal{O}_{  \bar{d}_{\rm R} d_{\rm L} \bar{e}_{\rm R} e_{\rm R}    D}(0,0)
		
		& i \left(D^{\mu } \overline{e}_{\rm R}{}_{s} d_{\rm L}{}_{ra}\right) \left(\overline{d}_{\rm R}{}_{p}^{a} \gamma _{\mu } e_{\rm R}{}_{t}\right)
		
	\end{array}\\
	
	&\begin{array}{c|l}
		
		\mathcal{O}_{  \bar{d}_{\rm R} d_{\rm L} \nu _{\rm L} \bar{\nu }_{\rm L}    D}(0,0)
		
		& i \left(d_{\rm L}{}_{ra} C D^{\mu } \nu _{\rm L}{}_{s}\right) \left(\overline{d}_{\rm R}{}_{p}^{a} \gamma _{\mu } C \overline{\nu}_{\rm L}{}_{t}\right)
		
	\end{array}\\
	
	&\begin{array}{c|l}
		
		\mathcal{O}_{  d_{\rm L} \bar{d}_{\rm L}  \bar{e}_{\rm R} e_{\rm L}   D}(0,0)
		
		& i \left(\overline{d}_{\rm L}{}_{t}^{a} \gamma _{\mu } d_{\rm L}{}_{pa}\right) \left(\overline{e}_{\rm R}{}_{r} D^{\mu } e_{\rm L}{}_{s}\right)
		
	\end{array}\\
	
	&\begin{array}{c|l}
		
		\mathcal{O}_{  d_{\rm L} \bar{e}_{\rm R} \nu _{\rm L} \bar{u}_{\rm L}    D}(0,0)
		
		& i \left(\overline{e}_{\rm R}{}_{r} D^{\mu } \nu _{\rm L}{}_{s}\right) \left(\overline{u}_{\rm L}{}_{t}^{a} \gamma _{\mu } d_{\rm L}{}_{pa}\right)
		
	\end{array}\\
	
	&\begin{array}{c|l}
		
		\mathcal{O}_{  d_{\rm L} \bar{d}_{\rm L}  \nu _{\rm L}^2   D}(0,2)
		
		&\mathcal{Y}\left[\tiny{\young(r,s)}\right]i \left(\nu _{\rm L}{}_{r} C D^{\mu } \nu _{\rm L}{}_{s}\right) \left(\overline{d}_{\rm L}{}_{t}^{a} \gamma _{\mu } d_{\rm L}{}_{pa}\right)
		
	\end{array}\\
	
	&\begin{array}{c|l}
		
		\mathcal{O}_{  \bar{d}_{\rm R}^3 e_{\rm R}    D}(-1,1)
		
		&\mathcal{Y}\left[\tiny{\young(prs)}\right]i \epsilon _{abc} \left(\overline{d}_{\rm R}{}_{p}^{a} \gamma _{\mu } e_{\rm R}{}_{t}\right) \left(\overline{d}_{\rm R}{}_{r}^{b} C D^{\mu } \overline{d}_{\rm R}{}_{s}^{c}\right)
		
	\end{array}\\
	
	&\begin{array}{c|l}
		
		\mathcal{O}_{  \bar{d}_{\rm R}^2 \bar{d}_{\rm L}  e_{\rm L}   D}(-1,1)
		
		&\mathcal{Y}\left[\tiny{\young(pr)}\right]i \epsilon _{abc} \left(\overline{d}_{\rm R}{}_{r}^{b} D^{\mu } e_{\rm L}{}_{s}\right) \left(\overline{d}_{\rm R}{}_{p}^{a} \gamma _{\mu } C \overline{d}_{\rm L}{}_{t}^{c}\right)
		
	\end{array}\\
	
	&\begin{array}{c|l}
		
		\mathcal{O}_{  \bar{d}_{\rm R}^2 \nu _{\rm L} \bar{u}_{\rm L}    D}(-1,1)
		
		&\mathcal{Y}\left[\tiny{\young(pr)}\right]i \epsilon _{abc} \left(\overline{d}_{\rm R}{}_{r}^{b} D^{\mu } \nu _{\rm L}{}_{s}\right) \left(\overline{d}_{\rm R}{}_{p}^{a} \gamma _{\mu } C \overline{u}_{\rm L}{}_{t}^{c}\right)
		
	\end{array}\\
	
	&\begin{array}{c|l}
		
		\mathcal{O}_{  \bar{d}_{\rm R} d_{\rm R}  \bar{e}_{\rm R} e_{\rm L}   D}(0,0)
		
		& i \left(\overline{d}_{\rm R}{}_{p}^{a} \gamma _{\mu } d_{\rm R}{}_{ta}\right) \left(\overline{e}_{\rm R}{}_{r} D^{\mu } e_{\rm L}{}_{s}\right)
		
	\end{array}\\
	
	&\begin{array}{c|l}
		
		\mathcal{O}_{  \bar{d}_{\rm R} e_{\rm L} \nu _{\rm L} u_{\rm R}    D}(0,2)
		
		& i \left(e_{\rm L}{}_{r} C D^{\mu } \nu _{\rm L}{}_{s}\right) \left(\overline{d}_{\rm R}{}_{p}^{a} \gamma _{\mu } u_{\rm R}{}_{ta}\right)
		
	\end{array}%
	
\end{align}

\subsection{Lists of the Dim-8 Operators}

\subsubsection{Classes involving Bosons only}

\indent \indent \underline{Class $F_{\rm{L}}{}^4      $}: 4 types

\begin{align}
	
	&%
	
\end{align}

\subsubsection{Classes involving Two-fermions}

\indent \indent \underline{Class $F_{\rm{L}}{}^2 \psi  \psi^\dagger   D$}: 8 types

\begin{align}
	
	&%
	
\end{align}

\subsubsection{Classes involving Four-fermions}

\indent \indent \underline{Class $  \psi {}^4   D^2$}: 18 types

\begin{align}
	
	&%
	
\end{align}

\subsection{Lists of the Dim-9 Operators}

\subsubsection{Classes involving Two-fermions}

\indent \indent \underline{Class $F_{\rm{L}}{}^2 \psi {}^2   D^2$}: 10 types

\begin{align}
	
	&%
	
\end{align}

\subsubsection{Classes involving Four-fermions}

\indent \indent \underline{Class $  \psi {}^3 \psi^\dagger   D^3$}: 52 types

\begin{align}
	
	&%
	
\end{align}

\subsubsection{Classes involving Six-fermions: $  \psi {}^6    $}

There are 58 complex types in this class.

\paragraph{$(\Delta B,\Delta L)=(0,0)$}
\begin{align}
    &%

\end{align}

\paragraph{$(\Delta B,\Delta L)=(0,\pm2)$}
\begin{align}
    &%

\end{align}

\paragraph{$(\Delta B,\Delta L)=(0,\pm4)$}
\begin{align}
    &%

\end{align}

\paragraph{$(\Delta B,\Delta L)=(0,\pm6)$}
\begin{align}
	&%

\end{align}

\paragraph{$(\Delta B,\Delta L)=(\pm1,\pm1)$}
\begin{align}
    &%

\end{align}

\paragraph{$(\Delta B,\Delta L)=(\pm1,\mp1)$}
\begin{align}
    &%

\end{align}

\paragraph{$(\Delta B,\Delta L)=(\pm1,\pm3)$}
\begin{align}
    &%

\end{align}

\paragraph{$(\Delta B,\Delta L)=(\pm1,\mp3)$}
\begin{align}
    &%

\end{align}

\paragraph{$(\Delta B,\Delta L)=(\pm2,0)$}
\begin{align}
    &%

\end{align}

\subsubsection{Classes involving Six-fermions: $  \psi {}^4 \psi^\dagger{}^2    $}

There are 410 complex types in this class.

\paragraph{$(\Delta B,\Delta L)=(0,0)$}
\begin{align}
    &%

\end{align}

\paragraph{$(\Delta B,\Delta L)=(0,\pm2)$}
\begin{align}
	&%

\end{align}

\paragraph{$(\Delta B,\Delta L)=(0,\pm4)$}
\begin{align}
	&%

\end{align}

\paragraph{$(\Delta B,\Delta L)=(\pm1,\pm1)$}

\begin{align}
	
	&%

\end{align}

\paragraph{$(\Delta B,\Delta L)=(\pm1,\mp1)$}

\begin{align}
	
	&%
	
\end{align}

\paragraph{$(\Delta B,\Delta L)=(\pm1,\pm3)$}

\begin{align}
	
	&%
	
\end{align}

\paragraph{$(\Delta B,\Delta L)=(\pm1,\mp3)$}

\begin{align}
	
	&%
	
\end{align}

\paragraph{$(\Delta B,\Delta L)=(\pm2,0)$}

\begin{align}
	
	&%
	
\end{align}

\section{Conclusion}\label{sec:con}
In this paper, we proved the validity of the on-shell method of operator enumeration~\cite{Li:2020gnx,Li:2020xlh} for effective field theories including massive fermions. While the intuition is that fermion masses do not alter the possible form of operators, we also give a rigorous proof from the amplitude perspective.
Hence we could apply the method to obtain the full operator basis at dimension 5 through 9 for LEFT, the effective field theory below the weak scale in the Standard Model where all fermions are massive. The completeness and independence of the basis are guaranteed by the method without subjective judgements.

At mass dimension 5, 6, 7, 8, 9, we conclude that there are 10, 118, 150, 756, 3686 independent operators respectively for one generation of fermions, and 76, 5875, 5118, 35058, 704584 flavor specified independent operators for three generation of fermions. The results of dimension 5 to 7 are consistent with previous existing results, with flavor symmetry specified. The total number of operators, as well as the number of operators with various lepton number violation $|\Delta L|$, are summarized for each dimension in table~\ref{tab:sumdim5678} and table~\ref{tab:sumdim9}. In tables~\ref{tab:BLstat5},~\ref{tab:BLstat6},~\ref{tab:BLstat7},~\ref{tab:BLstat8},~\ref{tab:BLstat9}, we list the number of types (operators) for each baryon-number and lepton-number violation pattern with different kinds of flavors involved at different scales, ranging from the MeV scale to the bottom quark mass scale.

\begin{table}
	\begin{align*}
		\begin{array}{|c|c|c|c|c|}
			\hline
			\multicolumn{5}{|c|}{\text{Dim-5 operators}}\\
			\hline
			\multirow{2}*{\diagbox{$(\Delta B, \Delta L)$}{$(n_u, n_d, n_e, n_\nu)$}} & \multirow{2}*{(2,3,3,3)} & \multirow{2}*{(2,2,3,3)} & \multirow{2}*{(1,2,2,3)} & \multirow{2}*{(1,1,1,3)} \\
            &&&&\\
            \hline
			(0,0) & 10 \, (70) & 10 \, (50) & 10 \, (28) & 10 \, (10) \\
			\hline
			(0,\pm2) & 2 \, (6) & 2 \, (6) & 2 \, (6) & 2 \, (6) \\
			\hline
		\end{array}
	\end{align*}
	\caption{The number of types (operators) of each baryon-number and lepton-number violation pattern with various flavors involved at dimension 5. $n_u$, $n_d$, $n_e$ and $n_\nu$ represent active number of up-type quarks, down-type quarks, charged leptons and neutrinos respectively. For $n_u=2$ and $n_u=1$, the corresponding active up-type quarks are $(c,u)$ and $(u)$ respectively, For $n_d=3,2,1$, active down-type quarks are $(b,s,d)$, $(s,d)$ and $(d)$ respectively, $n_e=3,2,1$ corresponds active charged letpons $(\tau,\mu,e)$, $(\mu,e)$ and $(e)$. The number of active neutrino are always set to 3.  The only type of $(0,\pm2)$ at dimension 5 is $F_L \nu_L^2$, so the numbers of operators with different flavor numbers stay the same.
	}
	\label{tab:BLstat5}
\end{table}

\begin{table}
	\begin{align*}
		\begin{array}{|c|c|c|c|c|}
			\hline
			\multicolumn{5}{|c|}{\text{Dim-6 operators}}\\
			\hline
			\multirow{2}*{\diagbox{$(\Delta B, \Delta L)$}{$(n_u, n_d, n_e, n_\nu)$}} & \multirow{2}*{(2,3,3,3)} & \multirow{2}*{(2,2,3,3)} & \multirow{2}*{(1,2,2,3)} & \multirow{2}*{(1,1,1,3)} \\
            &&&&\\
            \hline
			(0,0) & 56 \, (3631) & 56 \, (2171) & 56 \, (754) & 56 \, (183) \\
			\hline
			(0,\pm2) & 20 \, (1200) & 20 \, (870) & 20 \, (390) & 20 \, (120) \\
			\hline
			(0,\pm4) & 2 \, (12) & 2 \, (12) & 2 \, (12) & 2 \, (12) \\
			\hline
			(\pm1,\pm1) & 18 \, (576) & 18 \, (324) & 14 \, (86) & 12 \, (20) \\
			\hline
			(\pm1,\mp1) & 14 \, (456) & 14 \, (156) & 14 \, (86) & 4 \, (12) \\
			\hline
		\end{array}
	\end{align*}
	\caption{The number of types (operators) of each baryon-number and lepton-number violation pattern with various flavor numbers at dimension 6. $n_u$, $n_d$, $n_e$ and $n_\nu$ represent active number of up-type quarks, down-type quarks, charged leptons and neutrinos respectively. For $n_u=2$ and $n_u=1$, the corresponding active up-type quarks are $(c,u)$ and $(u)$ respectively, For $n_d=3,2,1$, active down-type quarks are $(b,s,d)$, $(s,d)$ and $(d)$ respectively, $n_e=3,2,1$ corresponds active charged letpons $(\tau,\mu,e)$, $(\mu,e)$ and $(e)$.
	}
	\label{tab:BLstat6}
\end{table}

\begin{table}
	\begin{align*}
		\begin{array}{|c|c|c|c|c|}
			\hline
			\multicolumn{5}{|c|}{\text{Dim-7 operators}}\\
			\hline
			\multirow{2}*{\diagbox{$(\Delta B, \Delta L)$}{$(n_u, n_d, n_e, n_\nu)$}} & \multirow{2}*{(2,3,3,3)} & \multirow{2}*{(2,2,3,3)} & \multirow{2}*{(1,2,2,3)} & \multirow{2}*{(1,1,1,3)} \\
            &&&&\\
			\hline
			(0,0) & 82 \, (3168) & 82 \, (1894) & 82 \, (698) & 78 \, (166) \\
			\hline
			(0,\pm2) & 30 \, (750) & 30 \, (546) & 30 \, (258) & 30 \, (114) \\
			\hline
			(\pm1,\pm1) & 18 \, (612) & 18 \, (360) & 18 \, (108) & 18 \, (30) \\
			\hline
			(\pm1,\mp1) & 14 \, (588) & 14 \, (240) & 14 \, (140) & 14 \, (26) \\
			\hline
		\end{array}
	\end{align*}
	\caption{The number of types (operators) of each baryon-number and lepton-number violation pattern with various flavor numbers at dimension 7. $n_u$, $n_d$, $n_e$ and $n_\nu$ represent active number of up-type quarks, down-type quarks, charged leptons and neutrinos respectively. For $n_u=2$ and $n_u=1$, the corresponding active up-type quarks are $(c,u)$ and $(u)$ respectively, For $n_d=3,2,1$, active down-type quarks are $(b,s,d)$, $(s,d)$ and $(d)$ respectively, $n_e=3,2,1$ corresponds active charged letpons $(\tau,\mu,e)$, $(\mu,e)$ and $(e)$.
	}
	\label{tab:BLstat7}
\end{table}

\begin{table}
	\begin{align*}
		\begin{array}{|c|c|c|c|c|}
			\hline
			\multicolumn{5}{|c|}{\text{Dim-8 operators}}\\
			\hline
			\multirow{2}*{\diagbox{$(\Delta B, \Delta L)$}{$(n_u, n_d, n_e, n_\nu)$}} & \multirow{2}*{(2,3,3,3)} & \multirow{2}*{(2,2,3,3)} & \multirow{2}*{(1,2,2,3)} & \multirow{2}*{(1,1,1,3)} \\
            &&&&\\
			\hline
			(0,0) & 281 \, (21144) & 281 \, (11872) & 277 \, (4140) & 267 \, (954) \\
			\hline
			(0,\pm2) & 78 \, (5442) & 78 \, (3810) & 78 \, (1668) & 78 \, (516) \\
			\hline
			(0,\pm4) & 4 \, (48) & 4 \, (48) & 4 \, (48) & 4 \, (48) \\
			\hline
			(\pm1,\pm1) & 78 \, (4536) & 78 \, (2592) & 74 \, (720) & 72 \, (180) \\
			\hline
			(\pm1,\mp1) & 58 \, (3888) & 58 \, (1440) & 58 \, (816) & 48 \, (132) \\
			\hline
		\end{array}
	\end{align*}
	\caption{The number of types (operators) of each baryon-number and lepton-number violation pattern with various flavor numbers at dimension 8. $n_u$, $n_d$, $n_e$ and $n_\nu$ represent active number of up-type quarks, down-type quarks, charged leptons and neutrinos respectively. For $n_u=2$ and $n_u=1$, the corresponding active up-type quarks are $(c,u)$ and $(u)$ respectively, For $n_d=3,2,1$, active down-type quarks are $(b,s,d)$, $(s,d)$ and $(d)$ respectively, $n_e=3,2,1$ corresponds active charged letpons $(\tau,\mu,e)$, $(\mu,e)$ and $(e)$.
	}
	\label{tab:BLstat8}
\end{table}

\begin{table}
	\begin{align*}
		\begin{array}{|c|c|c|c|c|}
			\hline
			\multicolumn{5}{|c|}{\text{Dim-9 operators}}\\
			\hline
			\multirow{2}*{\diagbox{$(\Delta B, \Delta L)$}{$(n_u, n_d, n_e, n_\nu)$}} & \multirow{2}*{(2,3,3,3)} & \multirow{2}*{(2,2,3,3)} & \multirow{2}*{(1,2,2,3)} & \multirow{2}*{(1,1,1,3)} \\
            &&&&\\
			\hline
			(0,0) & 640 \, (318062) & 640 \, (144230) & 640 \, (33046) & 634 \, (4546) \\
			\hline
			(0,\pm2) & 334 \, (143808) & 334 \, (75996) & 334 \, (19964) & 334 \, (3718) \\
			\hline
			(0,\pm4) & 20 \, (2136) & 20 \, (1494) & 20 \, (598) & 20 \, (172) \\
			\hline
			(0,\pm6) & 2 \, (2) & 2 \, (2) & 2 \, (2) & 2 \, (2) \\
			\hline
			(\pm1,\pm1) & 290 \, (121590) & 290 \, (52344) & 280 \, (9646) & 266 \, (1294) \\
			\hline
			(\pm1,\mp1) & 230 \, (106494) & 230 \, (33816) & 228 \, (10330) & 194 \, (1010) \\
			\hline
			(\pm1,\pm3) & 30 \, (4746) & 30 \, (3360) & 22 \, (498) & 16 \, (118) \\
			\hline
			(\pm1,\mp3) & 14 \, (2730) & 14 \, (840) & 14 \, (510) & 8 \, (46) \\
			\hline
			(\pm2,0) & 16 \, (5016) & 16 \, (1064) & 16 \, (288) & 16 \, (28) \\
			\hline
		\end{array}
	\end{align*}
	\caption{The number of types (operators) of each baryon-number and lepton-number violation pattern with various flavor numbers at dimension 9. $n_u$, $n_d$, $n_e$ and $n_\nu$ represent active number of up-type quarks, down-type quarks, charged leptons and neutrinos respectively. For $n_u=2$ and $n_u=1$, the corresponding active up-type quarks are $(c,u)$ and $(u)$ respectively, For $n_d=3,2,1$, active down-type quarks are $(b,s,d)$, $(s,d)$ and $(d)$ respectively, $n_e=3,2,1$ corresponds active charged letpons $(\tau,\mu,e)$, $(\mu,e)$ and $(e)$.
	}
	\label{tab:BLstat9}
\end{table}

The low energy effective field theory has profound guiding significance, and a complete operator basis of it at high dimensions provides various phenomenological applications, especially for the exotic processes with baryon or lepton number violations. With the complete set of operator basis, it would be very interesting to see how these higher dimensional operators are applied to various exotic processes and how the Wilson coefficients are constrained from current experimental data.

\section*{Acknowledgements}
We thank Christopher Murphy point out an arithmetic error in counting numbers of  the  $F_L{}^2 F_R{}^2$ operators, although we list all the explicit operators (\ref{cl:FL2FR2f})-(\ref{cl:FL2FR2l}) correctly which has been written in a systematical way.
H.L.L. Z.R. and J.H.Y. were supported by the National Science Foundation of China (NSFC) under Grants No. 11875003 and No. 12022514.  
M.L.X. is supported by the National Natural Science Foundation of China (NSFC) under grant No.2019M650856 and the 2019 International Postdoctoral Exchange Fellowship Program.
JHY was also supported by the National Science Foundation of China (NSFC) under Grants No. 11947302.

\bibliographystyle{JHEP}
\bibliography{LEFTref}

\end{document}